\documentclass[pdflatex,sn-basic]{sn-jnl}
\usepackage{multirow}
\usepackage{longtable}
\usepackage{afterpage}
\usepackage[margins]{trackchanges}

\jyear{2023}%

\theoremstyle{thmstyleone}%
\theoremstyle{thmstyletwo}%

\theoremstyle{thmstylethree}%

\renewcommand{\remove}[1]{}

\newcommand{\rev}[1]{\textcolor{black}{#1}}

\begin{document}

\addeditor{Wei Wang}
\title[Adaptive user interfaces in systems targeting chronic disease...]{Adaptive user interfaces in systems targeting chronic disease:
a systematic literature review}


\author*[1]{\fnm{Wei} \sur{Wang}}\email{wei.wang5@monash.edu}

\author[1,2]{\fnm{Hourieh} \sur{Khalajzadeh}}\email{hkhalajzadeh@deakin.edu.au}

\author[1]{\fnm{John} \sur{Grundy}}\email{john.grundy@monash.edu}
\author[1]{\fnm{Anuradha} \sur{Madugalla}}\email{anuradha.madugalla@monash.edu}
\author[1,3]{\fnm{Jennifer} \sur{McIntosh}}\email{jennifer.mcintosh@unimelb.edu.au}
\author[1]{\fnm{Humphrey O.} \sur{Obie}}\email{humphrey.obie@monash.edu}

\affil*[1]{\orgdiv{Department of Software Systems and Cybersecurity}, \orgname{Monash University}, \orgaddress{\city{Melbourne}, \country{Australia}}}

\affil[2]{\orgdiv{School of Information Technology}, \orgname{Deakin University}, \orgaddress{\city{Melbourne}, \country{Australia}}}

\affil[3]{\orgdiv{School of Population and Global Health}, \orgname{University of Melbourne}, \orgaddress{\city{Melbourne}, \country{Australia}}}


\abstract{eHealth technologies have been increasingly used to foster proactive self-management skills for patients with chronic diseases. However, it is challenging to provide each user with their desired support due to the dynamic and diverse nature of the chronic disease and its impact on users. Many such eHealth applications support aspects of `adaptive user interfaces' -- interfaces that change or can be changed to accommodate the user and usage context differences. To identify the state-of-art in adaptive user interfaces in the field of chronic diseases, we systematically located and analysed 48 key studies in the literature with the aim of categorising the key approaches used to date and identifying limitations, gaps and trends in research. Our data synthesis is based on the data sources used for interface adaptation, the data collection techniques used to extract the data, the adaptive mechanisms used to process the data and the adaptive elements generated at the interface. The findings of this review will aid researchers and developers in understanding where adaptive user interface approaches can be applied and necessary considerations for employing adaptive user interfaces to different chronic disease-related eHealth applications. }




\maketitle

\section{Introduction}
Chronic diseases, such as asthma, cardiac disease, and diabetes have become some of the biggest challenges facing the healthcare system \citep{Who}. As reported by the World Health Organisation (WHO), chronic diseases account for 74\% of fatalities each year \citep{Who}. Given an increasing allocation of healthcare resources to chronic disease management, the medical care paradigm is shifting from hospital-based reactive treatment to long-term self-management \citep{talboom2018chronic,di2019chronic}. eHealth technologies have been increasingly used to foster such proactive self-management skills and to help prevent the development of secondary complications through mechanisms such as digital education, self-monitoring, and feedback \citep{free2013effectiveness}. Numerous studies have demonstrated the value of eHealth technology in managing chronic diseases \citep{pare2007systematic}. These technologies often focus on addressing particular chronic diseases and monitoring essential physiological indicators. 

However, people with chronic diseases exhibit significant diversity in their symptoms, condition severity and diverse human characteristics \citep{harvey2012future}. \rev{These individuals are often of different ages, physical and mental challenges, cultural diversities and educational attainment, all of which influence the utilisation of eHealth systems \citep{luy2021toolkit}. An additional layer of complexity arises from the inherently prolonged nature of chronic diseases \citep{Who,harvey2012future}, coupled with their susceptibility to various secondary complications and comorbidities} \citep{gregor2002designing,di2019chronic}. To cope with the variability, some eHealth applications take the context of usage into account  \citep{floch2018user,setiawan2019adaptive}. However, some of them only apply simple predefined rule sets, or fail to take into account the user's unique traits and behavioural characteristics \citep{grua2020reference,mclean2011telehealthcare}.  In this context, an \textbf{Adaptive User Interface (AUI)} may provide a viable solution to contextual variability \citep{Akiki2014} and an effective instrument for keeping users continuously engaged and actively involved, eventually leading to better physical and mental health outcomes \citep{grua2020reference,floch2018user}. \cite{mctear1993user} defines an AUI as  \emph{“a software artefact that improves its ability to interact with a user by constructing a user model based on partial experience with that user”}. A key goal of an AUI is to incorporate individual user perceptions, resulting in more effective system use with reduced error and frustration \citep{Vasilyeva2005,weld2003automatically,luy2021toolkit}. This study presents a \textbf{Systematic Literature Review (SLR)} focusing on the implementation of AUIs in applications that target chronic diseases or \textbf{risk factors associated with chronic diseases (RFCD)}.

 Some literature reviews have addressed the utilisation of  AUI for eHealth interventions \citep{Aranha2021, Goncalves2019, Palomares-Pecho2021, Robinson2020, OpdenAkker2014, Hachey2012, Sanchez2018}. However, \rev{several reviews exhibit limitations by predominantly concentrating on adapting the \textbf{user interface (UI)} to accommodate one specific factor, such as a user's emotional state while neglecting other crucial considerations} \citep{Aranha2021, Robinson2020}. Other reviews have examined specific application domains, for example, \textbf{Ambient Assisted Living (AAL)} applications \citep{Sanchez2018}, physical activity coaching applications \citep{OpdenAkker2014}, applications using gamification \citep{Robinson2020} and rehabilitation applications \citep{Palomares-Pecho2021}. Such limitations restrict the breadth of coverage and, consequently, the applicability of findings across other software domains.
 
 In comparison, this SLR explores the depth and breadth of evidence in the realm of AUIs within the chronic disease domain, employing a thorough and systematic analytical approach. In addition to discussing common adaptation components, this SLR goes further by discussing \textbf{1)} applications targeting chronic diseases or RFCD (e.g., unhealthy diet, harmful use of alcohol, physical inactivity and obesity), \textbf{2)} \rev{a diverse target audience encompassing healthcare professionals, individuals with chronic diseases, and those seeking preventive measures against chronic diseases, \textbf{3)} a comprehensive examination of UI adaptation properties (e.g., responsible parties for adaptation and adapted UI elements)}, and \textbf{4)} the inclusion of all types of software applications (e.g., web, mobile, desktop tablet applications). 

The objective of this SLR is to \textbf{\emph{provide a holistic view of the existing literature on the use of AUIs in applications that target chronic diseases or RFCD, while unveiling patterns and trends among various AUI solutions}}. We organise our work around five major \textbf{Research Questions (RQs)} that can be directly linked to the objective of this SLR. To answer our RQs, we systematically identified and rigorously reviewed 48 relevant papers and synthesised the data extracted from those papers. \rev{Our effort culminated in a taxonomy delineating important techniques and strategies characterised by distinct types of proposed AUI solutions.} The key contributions of this work include: 
\begin{itemize}
    \item a classification encompassing adaptation data sources, data collection techniques, adaptive strategies, adaptation actors and adaptive elements for different AUI solutions;
    \item insights into the link between different adaptation proprieties and the connections between adaptation proprieties and specific types of applications;
    \item providing the empirical \textbf{Software Engineering (SE)} community with useful insights about the AUI evaluation; and
    \item a list of key issues to guide future research endeavours aimed at facilitating the development and utilisation of AUIs within the chronic disease domain.
\end{itemize}

The remaining sections of this paper are structured as follows. We introduce the AUI concepts utilised in this study (Section \ref{subsec:adaptivuser}), followed by an exploration of eHealth technology's application in the chronic disease domain (Section \ref{subsec:eHealth}). Following this, we delve into the process of planning and executing an SLR in Section \ref{sec:method}. The resulting analysis for each RQ is presented in Sections \ref{sec:overview}, \ref{sec:RQ1}, \ref{sec:RQ2}, \ref{sec:RQ3}, \ref{sec:RQ4} and \ref{sec:RQ5}. Section \ref{sec:discussion} reports a discussion of the synthesis of the findings. The threats to validity are discussed in Section \ref{sec:threat}. In conclusion, we present our final remarks in Section \ref{sec:conclusion}.

\section{Background and Motivation }
\subsection{Adaptive user interfaces}\label{subsec:adaptivuser}
The immediate point of contact between users and their software resides within the UI. Therefore, it is crucial that users can easily communicate with the UI and interpret UI outputs \citep{Vogt2010}. The notion of making UI design work for a broad spectrum of individuals is promoted by multiple UI development methodologies, such as \emph{Universal Design} \citep{mace1991accessible} and \emph{Design for All} \citep{stephanidis1997towards}. However, a UI is not independent of its \emph{context of use} which is defined in terms of the user, platform, and environment \citep{Calvary2003}. Due to the diverse nature of users, 
achieving universal solutions necessitates individuals to adjust their behaviour and problem-solving approaches to effectively engage with the UI \citep{Norcio1989}. Additionally, the high prevalence of smartphone use offers new and flexible methods for interacting with information. This includes leveraging diverse wireless sensors that provide opportunities for collecting contextual data, such as physiological parameters \citep{Vogt2010}. As a result, a UI initially designed for a specific fixed context may no longer be sufficient. The \emph{adaptation of the UI} emerges as a plausible solution to address contextual variations, enabling seamless alignment within the context of use.

Depending on the allocation of adaptation responsibility, there are three key types of adaptation: \textbf{a)} \textit{Adaptability} (also known as \textit{manual systems}): where end users are granted explicit control to modify specific UI elements according to their needs \citep{Akiki2014,luy2021toolkit}; \textbf{b)} \textit{Adaptivity} (also known as \textit{automatic systems}): UI adjustments occur automatically in response to contextual changes \citep{Harman2014, Oreizy1999}; and \textbf{c)} \textit{Semi-automatic systems}: this approach integrates both \emph{adaptability} and \emph{adaptivity}, involving collaborative adaptation efforts between the system and end-users \citep{Aranha2021,mukhiya2020adaptive, Palomares-Pecho2021}. Besides the three main types of adaptation, several particular forms of adaptation can be achieved with manual, automatic or semi-automatic systems. For example, \textit{personalisation}, also referred to as \textit{customisation} or \textit{tailoring}, is a particular form of adaptation that typically targets the UI content \citep{abrahao2021model, Akiki2014, luy2021toolkit}. Furthermore, certain forms of adaptation, known as \textit{multi-targeting UI}, or \emph{multi-platform UI}, enable seamless functioning on different platforms and devices \citep{Calvary2003,grundy2002developing}, primarily concentrating on the technical aspects of UI adaptation. Our study includes \emph{manual}, \emph{automatic} and \emph{semi-automated systems} and particular forms of adaptations (e.g., \emph{personalisation} and \emph{multi-targeting UI}).

AUIs have been widely applied across various domains. In the field of \textit{education}, AUIs have found applications in learning systems to offer customised learning experiences that cater to individual variations \citep{kolekar2019rule}. \cite{graf2007providing} introduced a system that integrates adaptivity into courses by considering students' individual learning styles. Several studies have explored the generation of adaptive assessment questions based on the difficulty level and performance of students \citep{brusilovsky2005individualized,mangaroska2019elo}. Moreover, \cite{yang2013development} introduced an adaptive personalised presentation module that integrates various dimensions of personalised features, such as students' cognitive style and learning styles, to enhance the effectiveness of the learning experience. In the context of \textit{cultural heritage}, AUI techniques have been employed to design interactive interfaces and systems that adapt to the preferences, interests, and cultural backgrounds of users, providing personalised and immersive experiences \citep{christodoulou2019personalized,michalakis2022context}. \cite{trichopoulos2021augmented} developed a personalised digital storytelling system for cultural heritage sites. By leveraging context-aware and personalisation methods, the system delivers culturally relevant information based on user profiles, location, user movement and behaviour. 

Some studies highlight the application of AUIs in the \textit{transportation safety and security} domain. For example, \cite{hudlicka2002assessment} developed an adaptive framework that aimed to dynamically adjust the format and content of the system interface. The adaptation process considers factors such as the pilot's affective state, personality traits, and situation-specific beliefs that could potentially impact performance in a sweep mission task. In other research by \cite{nasoz2010affectively}, a novel approach was proposed to enhance driving safety through the use of multimedia technologies. This study emphasises the recognition and adaptation of a car to drivers' emotions using multi-modal intelligent car interfaces, with the objective of improving safety and optimising performance in critical transportation contexts. AUIs have also been employed in the \textit{workplace} to improve user efficiency and performance in dynamic working environments with evolving sensors and device integration. \cite{bongartz2012adaptive} utilised an application in the retail industry's distribution centre as an illustration of employing various modalities based on pre-defined adaptation rules.  

In our SLR, we aim to investigate the application of AUIs in managing and supporting individuals with chronic conditions or RFCD. As a result, studies outside the scope of chronic diseases related healthcare applications will not be considered in our research.

\subsection{eHealth applications for chronic disease management}
\label{subsec:eHealth}
Chronic diseases, often referred to as non-communicable diseases (NCDs) \citep{Who}, present a major challenge to healthcare, exacerbated in part by the continuous rise in the numbers of people afflicted by these conditions. Medical advancements have improved treatment outcomes, allowing many people to survive conditions that were previously fatal \citep{beaglehole2008improving}. Additionally, the ageing population further contributes to the growing burden of chronic diseases \citep{abowd1999towards}. The management of these persistent health conditions transcends the confines of biological parameters alone, with a growing emphasis on empowering patients to take an active role in self-management \citep{Who}. Empowering patients to self-manage chronic disease demands the acquisition of skills and techniques to manage their health \citep{abowd1999towards}. This has led to an increasing emphasis on developing technologies that can be applied to self-management. eHealth, for instance, encompasses a variety of technologies including computers, smartphones and wireless communications, providing avenues for patients to engage with their health \citep{talboom2018chronic}. For example, new technologies have been developed to facilitate medication adherence and improve self-tracking capabilities to aid in self-management \citep{pare2007systematic,hamine2015impact}. 

However, research indicates a paradox where individuals who could potentially benefit the most from eHealth solutions often exhibit lower usage rates \citep{han2010professional}. To increase the successful deployment of eHealth applications, especially for patients with chronic conditions, these applications should cater to diverse user groups. There are well-recognised challenges to doing this including: \textbf{Firstly}, chronic diseases are highly \emph{heterogeneous} in the way they affect patients (i.e., triggers, symptoms, severity) \citep{harvey2012future,deiss2006improved,audulv2013over}, leading to substantial variations in patients' self-management requirements \citep{hanlon2017telehealth}. \textbf{Secondly}, the design of the eHealth application should take into account the phases of chronic disease \emph{change over time} \citep{gregor2002designing,di2019chronic,audulv2013over}. Either deteriorating as per the natural history of the disease if left untreated or improving when given the right care \citep{lorig2003self}. In response to the evolving physical and mental change inherent with chronic disease over time, a self-management plan needs to be modifiable over time as well. For example, individuals with diabetes exhibit distinct needs during the early stages of diagnosis compared to when the condition has advanced into a chronic state. Furthermore, chronic diseases are often associated with other comorbidities including physical and/or psychopathological disorders \citep{di2019chronic}. For example, the trajectory of diabetes is associated with a plethora of potential health consequences, including vision loss, limb amputation, neuropathy, end-stage renal disease, cardiovascular disease, infections, and cognitive impairment \citep{vijan2015type}. This association not only results in increased use of healthcare services but also introduces a broader spectrum of user characteristics and functional necessities\citep{gregor2002designing}. \textbf{Thirdly}, the majority of chronic diseases are \emph{long-lasting} - generally lifelong \citep{Who,harvey2012future} - therefore, there is a need for eHealth technologies to keep users engaged and motivated over time. In addition to the diverse nature of chronic diseases, both health professionals and patients are individuals with diverse backgrounds, expertise and different demographic, psychological, and cognitive characteristics \citep{Vasilyeva2005,luy2021toolkit}. The diverse and dynamic nature of users with chronic disease highlights the necessity of having the AUI, which aims to improve the interaction between the user and UI by tailoring the UI to match users' prevailing goals and needs \citep{Norcio1989, Akiki2014}. Recent research has demonstrated the necessity of adaptability of eHealth technologies in facilitating chronic disease self-management \citep{floch2018user,grua2020reference}. By providing these adaptations, eHealth applications can enhance acceptance and motivation among users, thereby promoting increased usage and engagement  \citep{han2010professional,floch2018user}.

\subsection{Prior Surveys and Reviews}
The focus of our study is on the use of AUIs within the domain of chronic diseases. This section reviews related secondary research, which includes SLRs, systematic mapping reviews and surveys, that explore the utilisation of AUI. Notably, the extent and depth of analysis of AUIs within these studies exhibit variability. Prior to our work, there have been four existing SLR papers that discuss literature in a variety of areas related to AUI \citep{Aranha2021, Goncalves2019, Palomares-Pecho2021, Robinson2020}. \rev{\cite{Goncalves2019} conducted a SLR that explored intelligent UIs within software systems related to the \textit{Internet of Things} and smart cities.} However, that SLR is limited to discussing preliminary results and to date, the full paper with the final conclusions remains unpublished. Additionally, two other SLRs focused on one particular adaptation dimension, \emph{emotional state}, and its influence on the behaviour of UI \citep{Aranha2021, Robinson2020}. Another SLR by \cite{Palomares-Pecho2021} conducted a comprehensive examination of the literature on end-user adaptable technologies supporting rehabilitation. This is limited as it is primarily centred on applications customised by therapists, with minimal consideration for the role of patients.

As well as the SLRs mentioned above, two mapping reviews have also contributed to the context of this study \citep{Hachey2012, Sanchez2018}. \cite{Hachey2012} investigated web UI and its interplay with SE techniques. Although the authors extensively explored the technical dimensions of the semantic web field, their work does not delve into the exploration of user influence on the adaptation process. \cite{Sanchez2018} undertook a thorough mapping of the landscape of research in intelligent UI. While some of the literature aligns with the context of AUIs within the domain of chronic diseases, the scope of relevance is limited due to the emphasis on AAL. One existing survey conducted by \cite{OpdenAkker2014} offers a thorough overview of the tools and methods currently being applied for tailoring physical activity coaching applications. The survey also introduces a tailoring model encompassing seven distinct tailoring concepts within the realm of physical activity coaching. However, the primary purpose of this tailoring model is to illustrate how different tailoring concepts can be combined to adapt the motivational message rather than the UI adaptation.

\section{Research Methodology}\label{sec:method}
\begin{figure}[h]
  \centering
  \includegraphics[width=1.05\linewidth]{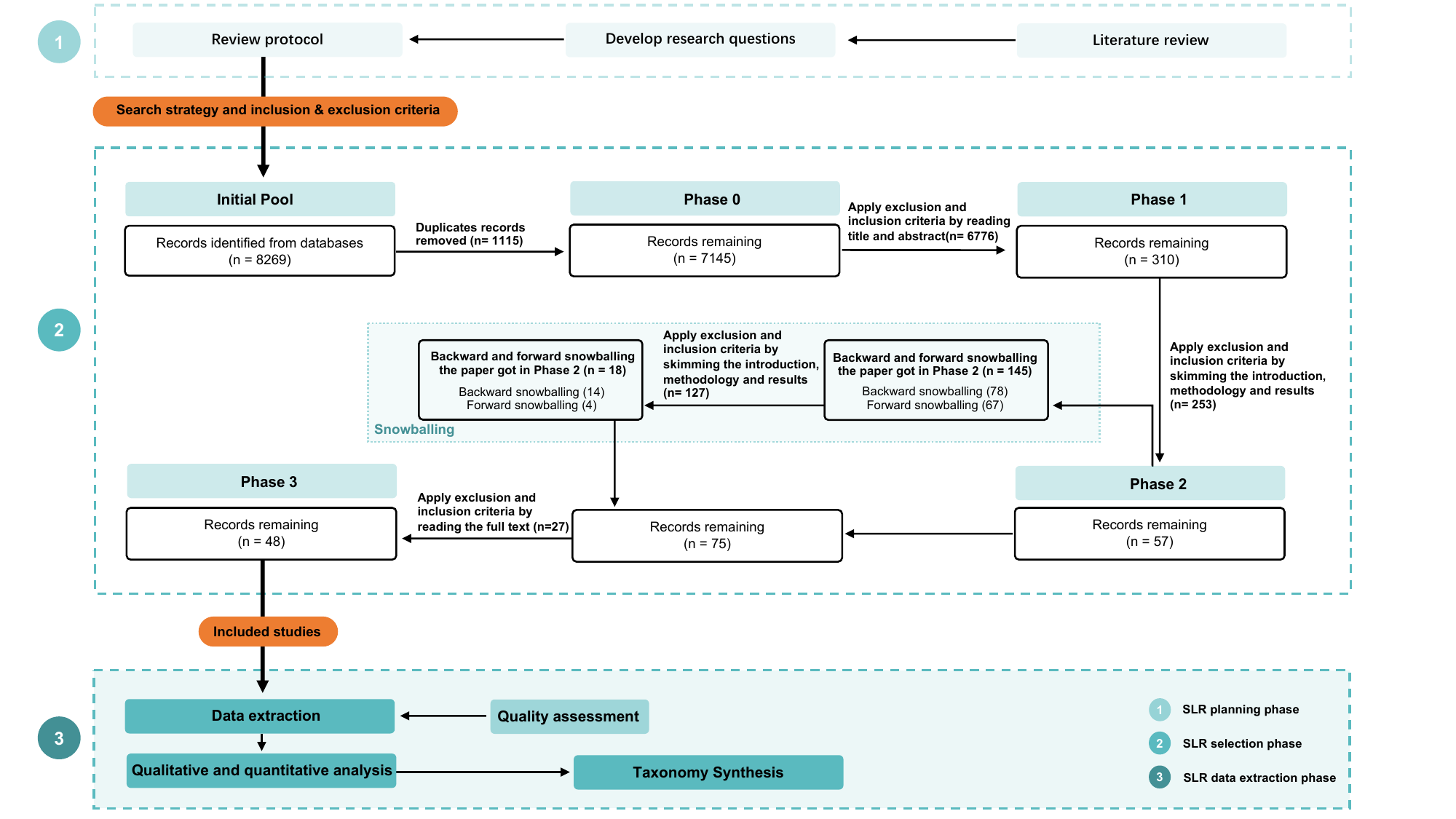}
  \caption{Stages of SLR process}
  \label{fig:flowchart}
\end{figure}

We followed the SLR guidelines and procedures in \citep{kitchenham2022segress} and the work of \citep{watson2022systematic} to uphold the integrity of our analysis and provide a reproducible method. Firstly, one author developed the SLR protocol which was reviewed by three authors to reduce bias. Our protocol identifies the key objectives of the review, the necessary background, RQs, inclusion and exclusion criteria, search strategy, data extraction, and analysis of gathered data. Figure \ref{fig:flowchart} depicts the process of our methodology, illustrating the three main steps of our review: \emph{SLR planning, selection and data extraction}. These three steps are covered in the following subsections with additional data in the supplementary material and appendix.

\subsection{Research Questions}
\label{sub:researchquestion}
 Our goal is to formulate RQs that logically guide the construction of a taxonomy of the surveyed research, while also tackling challenges associated with the design of AUIs within the domain of chronic diseases. The articulated RQs for this study are as follows:

\textbf{RQ\textsubscript{1}}: \rev{\textit{How are AUIs currently being used?} It is necessary to understand what has been accomplished so far in terms of AUIs for applications in the chronic disease domain.} This RQ aims to delve into several aspects, including the specific software platforms employed by researchers, the type of solutions that have been proposed to address pertinent health conditions and the identification of the specific user groups targeted by these proposed solutions.

\textbf{RQ\textsubscript{2}}: \textit{\rev{How is data being extracted, prepared, and used in the AUI?}} The basic components of an AUI are defined by the data presented to the corresponding application for adaptation purposes \citep{Norcio1989, Akiki2014, Calvary2003}. This RQ aims to obtain a taxonomy of the data used in the adaptation, how the data is retrieved, and how the data correlates to different application domains. Given the intricate nature of data selection and extraction, we dedicated two Sub-RQs to explore the use of data in AUIs:
\begin{itemize}
\item \textbf{RQ\textsubscript{2a}}: \textit{What types of data are collected to generate an AUI?} RQ\textsubscript{2a} investigates many forms of data utilised in AUIs. Given the large number of applications that currently gather data from users, it is critical to understand the sorts of data that are being used. 
\item \textbf{RQ\textsubscript{2b}}: \textit{How is data being extracted/collected?} RQ\textsubscript{2b} examines how data can be extracted. Furthermore, data capturing techniques are frequently reliant on the sort of data that the application seeks to extract which aids in discussing the relationship between data collection techniques and the data obtained. 
\end{itemize}

\textbf{RQ\textsubscript{3}}: \textit{What are the adaptive mechanisms used in generating the AUI?} The adaptive mechanisms typically consist of \emph{adaptive strategy} and \emph{adaptation actor}. This RQ aims to explore the decision-making processes that drive changes in the AUI. Considering the different aspects of the decision-making processes, RQ\textsubscript{3} is explored through two Sub-RQs:
\begin{itemize} 
\item \textbf{RQ\textsubscript{3a}}: \textit{What types of adaptive strategies are used to generate the AUI?} RQ\textsubscript{3a} explores the different adaptive strategies used to change the UI. We also look at common adaptive strategy pairings and how adaptive strategies are selected for various applications.
\item \textbf{RQ\textsubscript{3b}}: \textit{\rev{What roles do different actors play in AUI adaptation?}} RQ\textsubscript{3b} examines the different adaptation actors that trigger the adaptation process. We also investigate the interplay between the adaptation actors, adaptation strategies and corresponding applications.
\end{itemize}

\textbf{RQ\textsubscript{4}}: \textit{What are the adaptive elements employed in the AUI?} This RQ investigates specific adaptations demonstrated in different applications. We aim to construct a taxonomy outlining the range of adaptive elements utilised, frequent combinations of these elements, and how the adaptive elements are associated with different application domains. Our goal is to inspire researchers to explore the utilisation of adaptive elements in meaningful ways, effectively aligning specific adaptive elements with suitable applications.

\textbf{RQ\textsubscript{5}}: \rev{\textit{How is the AUI developed and evaluated?} This RQ investigates the design and evaluation approaches used by the included primary studies. Specifically, we examine the chosen evaluation approaches, indicators, and resulting evaluation outcomes presented in included primary studies. The findings of this RQ will aid the SE community in comprehending the factors within the evaluation process for AUI that have been inadequately described or overlooked. This understanding is crucial as it contributes to the challenges encountered in validating proposed solutions.}

\subsection{Search strategy}
Our search strategy aims to identify and collect all literature that complies with the inclusion and exclusion criteria detailed in Section \ref{sec:inclusion}. A mixed search strategy is adopted, incorporating both automatically searching through electronic databases and manually searching through conference and journal proceedings. 

\subsubsection{Data sources}
We surveyed and screened the search engines used in previous literature reviews in SE \citep{Maplesden2015,Shahin2014}. The list of electronic databases we eventually decided to search are: ACM Digital Library, IEEExplore, Springer link, ScienceDirect, Scopus and Medline. Wiley, Compendex, and Inspec were excluded due to their high overlap with other search engages. Both Compendex and Inspec overlap  with Scopus \citep{Maplesden2015}. Wiley is indexed by Scopus \citep{Wiley}. We faced certain difficulties when conducting the search. First, the SpringerLink search engine does not support title, abstract and keyword searches at the same time. We either needed to search for the full text of the article or the title only. The former yielded 61,722 papers, while the latter strategy, in contrast, only returned 4 papers. To address this issue, we followed the approach used in \citep{Maplesden2015} and included the top most relevant 2,000 papers returned by the full-text search. Another challenge is the limited number of search terms allowed in Science Direct, which demands splitting the search string into multiple search strings. Finally, digital libraries like ACM, IEEE, and Medline cannot simultaneously provide the ability to limit searches to multiple specific areas, e.g. title, abstract, and keywords combined, but Scopus can be used as a complement to other databases because it indexes the majority of SE articles and conferences \citep{Kitchenham2010}.

\subsubsection{Search terms}
PICOC criteria \citep{Kitchenham2007} were used to determine the search terms, shown in Table \ref{tab:searchtermcc}. Instead of searching for specific chronic conditions, we employed broad phrases from the health domain for two reasons: \textbf{1)} managing an excessive number of search terms could be challenging in certain databases (e.g., ScienceDirect), and \textbf{2)} exhaustively incorporating chronic disease search terms in automated searches is a formidable task. It is important to note that our study selection process adheres to the inclusion criteria to ensure that only studies relevant to chronic diseases are included. The chosen search terms have been modified and refined to fit each search engine since the chosen digital libraries have different constraints and their own unique search syntax. To ensure optimal results, multiple iterations of trial searches were conducted across six online databases. As part of this validation process, a set of 10 papers was randomly selected from each database, allowing us to verify the relevance and appropriateness of the compiled list of studies for our review.

\begin{table}[ht]
  \caption{Major search terms according to PICOC criteria}
  \footnotesize
  \centering
    \begin{tabular}{lp{95mm}}
    \toprule
    Concepts&Major search terms\\
    \midrule
   Population & (Application OR smartphone OR "smart phone" OR "cellular phone" OR "cell phone" OR cellphone OR software OR android OR ios OR windows OR tablet OR iPad) \\
    Intervention & (adapt* OR tailor* OR flexib* OR  personali?e* OR customi?e* OR context*-aware*) AND (GUI OR interface OR UI OR "user experience" OR UX OR usability)\\
    Context & (health OR healthcare OR ehealth OR mhealth OR disease? OR disorder? OR illness* OR patient?) \\
    \bottomrule
    \end{tabular}%
  \label{tab:searchtermcc}%
\end{table}%

\subsection{Inclusion and exclusion criteria}\label{sec:inclusion}
\rev{Some studies do not offer the necessary information to find the answer to our RQs. The inclusion and exclusion criteria we used are applied to all retrieved studies from databases and they are systematically applied at different stages of the selection process.}
\begin{itemize}
\item \textbf{I01}: Full-text papers published as journal or conference papers that focus on AUIs in eHealth applications targeting chronic disease.
\item \textbf{I02}: Entire papers are written in English and use academic literature references.
\item \textbf{I03}: The study must be available in full text and published in a renowned digital library.
\item \textbf{E01}: Grey literature, Workshop articles, posters, books, work-in-progress proposals, keynotes, editorial, secondary or review studied.
\item \textbf{E02}: Short papers less than four pages, irrelevant and low-quality studies that do not contain a considerable amount of information for AUI to extract.
\item \textbf{E03}: Extended or recent journal version is available (from same authors) on the same work. 
\item \textbf{E04}: Papers related to eHealth application using AUI that do not discuss chronic disease.
\item \textbf{E05}: Papers that solely describe the recommendation system or discuss the interface in Virtual Reality (VR) and Augmented Reality (AR).
\end{itemize}

\rev{The exclusion criterion E04 covers a wide range of chronic diseases, adhering to the chronic disease classification provided by the WHO \mbox{\citep{Who}}. Notably, Autism is regarded as a chronic disease in this SLR, aligning with current recommendations proposed by \mbox{\cite{davignon2018psychiatric}}.}

The exclusion criterion E05 pertains to papers that solely focus on either VR/AR or recommendation systems. While recommendation systems and AUI may exhibit some overlap in terms of personalisation \citep{Akiki2014,isinkaye2015recommendation}, they address distinct aspects of user interaction and customisation. Consequently, papers exclusively discussing the recommendation system are excluded. Furthermore, VR/AR user interfaces and normal user interfaces significantly differ in terms of user interaction and presentation. Consequently, to maintain the specific scope and objectives of our study, papers solely discussing VR and AR solutions are also excluded from our study.

\subsection{Study Selection}
Figure \ref{fig:flowchart} shows the number of studies retrieved at each stage of this SLR. The selection of primary research was performed using predetermined inclusion and exclusion criteria (see Section \ref{sec:inclusion}). From the first stage to the final screening, the essential records of the papers were kept in Excel spreadsheets and the Mendeley library. Separate sheets were kept in Excel to keep track of the selection decisions for each phase. By adopting these techniques, the consistency of the inclusion and exclusion criteria could be verified. Additional data for the selection process are given in supplementary material. The selection process is effectively divided into four phases:

\textbf{Phase 0}: We ran the search string on the six digital libraries and retrieved 7,145 papers after removing the duplicates.

\textbf{Phase 1}: Publications found during the initial search were assessed for their suitability based upon analysis of their title and abstract. Studies were then transferred to the next round of screening for further investigation if it was not possible to decide by reading the titles and abstract. At the end of this phase, 310 papers were selected.

\textbf{Phase 2}: Publications selected during Phase 1 went through a more thorough analysis (by skimming the introduction, methodology and results). 23 papers were chosen as a random sample and were reviewed by two co-authors. Two co-authors and the first author agreed on the study selection in over 75\% of the studies. Disagreement was easily resolved through discussion with the third author. As a result, 57 papers were later included. Then we applied both backward and forward snowballing techniques \citep{Wohlin2014} and found a total of 18 potentially relevant papers. This phase resulted in 75 papers.

\textbf{Phase 3}: Publications selected during Phase 2 were engaged in full-text screening. We excluded several papers beyond the planned scope and do not provide sufficient data to answer our RQs. At the end of Phase 3, we selected 48 papers for review.

\subsection{Study quality assessment}\label{subsec:qualityassess}
Each publication in the final set was assessed for its quality. The \textbf{quality assessment(QA)} procedure occurred at the same time as the extraction of relevant data to ensure that the findings of each individual study added substantial value to the SLR. A set of study quality assessment questions are listed in Table \ref{tab:qualityassessment} in Appendix \ref{app:qualityass}. These questions were adopted and adjusted from \citep{Kitchenham2007}. Each paper was then assigned a categorisation score ranging from 1 to 3, denoting low to high quality, based on responses to the QA questions.

\subsection{Data extraction and synthesis strategy}
To answer RQ\textsubscript{1}–RQ\textsubscript{5} and facilitate the data extraction process, we used the data extraction form in Table \ref{app:dataeextraction} in Appendix \ref{app:dataeextraction} to collect necessary information from the included studies. Additional data for the data extraction process are given in the supplementary material. In this study, the data extraction procedure was divided into three phases.

\textbf{Phase 1}: \rev{Data extraction employed a Google Form, refined over three iterations by extracting data from a single paper per database. Iterations focused on enhancing extraction question formulation and structural clarity for the form.}

\textbf{Phase 2}: After the questions and extraction form were finalised, the Google form was then sent to three co-authors for the same extraction for 6 papers. Then we did a comparison to check if there were any conflicting extracted information until all conflicts were resolved. Agreement about coded items before reaching consensus is quantified by using the \emph{percent agreement} \citep{hsu2003interrater}. It is important to highlight that the agreement is only assessed for the most critical data (e.g. adaptive strategy, adaptive elements, etc.). Our overall \emph{percent agreement} is 82\% with only a small number of outlier scores.

\textbf{Phase 3}: After further discussion and consensus on the disagreements, the first author, under the close supervision of the second and third authors, re-extracted the data from the previously reviewed studies as well as the remaining 42 studies. The reliability of our data extraction method was also checked during the extraction process. In the case of finding recurrent disagreement concerning certain data, we adjusted the coding instructions accordingly. All extracted data were kept in a spreadsheet, which allowed for quick reference while drafting the report.

The essential information, including demographic data, answers to RQs, and QA scores, was extracted and compiled into a data extraction sheet. Both qualitative and quantitative methods were employed for data synthesis. The quantitative analysis involved univariate and multivariate frequency distribution analysis, using tools such as Python and Microsoft Excel pivot tables for visualisation. Qualitative analysis followed an \emph{open coding methodology} consistent with \emph{constructivist grounded theory} \citep{charmaz2014constructing}, consisting of initial coding and focused coding to build a taxonomy on the use of AUI \mbox{\citep{stol2016grounded}}. The coding process was performed by multiple authors and resulted in the formation of the taxonomy, which is further detailed in Appendix \ref{app:datasynthesis}.

\section{Overview of included studies}\label{sec:overview}
In this section, we present a concise summary of the included studies, focusing on study demographics analysis. Table \ref{tab:searchresult} provides an overview of the primary studies with respect to their source of database. Notably, the highest percentage of relevant studies (10 studies, 21\%) was found in Scopus, whereas snowballing yielded the largest number of relevant studies (14 studies, 29\%). Table \ref{tab:listofstudies} in Appendix \ref{appe:includedstu} lists the 48 primary studies in our review.

{\footnotesize
\begin{table}[ht]
  \caption{Number of included studies for each study selection phase}
  \footnotesize
  \centering
    \begin{tabular}{p{22mm}p{12mm}p{12mm}p{12mm}p{12mm}p{25mm}}
    \toprule
    \textbf{Digital library}&\textbf{Phase 0}&\textbf{Phase 1}&\textbf{Phase 2}& \textbf{Phase 3}&\textbf{\% of all relevant studies}\\
    \midrule
    ACM&653&72&11&6&13\% \\
    IEEE&645&57&10&5&10\%\\
    Medline&1523&45&6&5&10\%\\
    ScienceDirect&1273&29&5&4&8\%\\
    Scopus&1147&35&15&10&21\%\\
    SpringerLink&1913&72&10&4&8\%\\
    Snowballing&0&0&18&14&29\%\\
    All libraries&7154&310&75&48&100\%\\
    \bottomrule
    \end{tabular}%
  \label{tab:searchresult}%
\end{table}}

From these results, we found that a significant amount of research comes from academia, i.e., 73\% of papers, followed by a small amount of research (17\%) from industry-academic collaborations. There are limited studies from government initiatives (4\%) and industry (6\%). To date, most applications with AUI seem to be confined to academia. Most included studies met QA criteria with clear objectives and contextual information. However, improvements are needed in describing the AUI, reporting outcome analysis, and providing comprehensive findings, limitations, and future work statements. The subsequent sections present an in-depth analysis and responses to our five RQs. 

\section{RQ\textsubscript{1}: How are AUIs currently being used?}\label{sec:RQ1}
\begin{figure}[ht]
\includegraphics[width=0.95\linewidth]{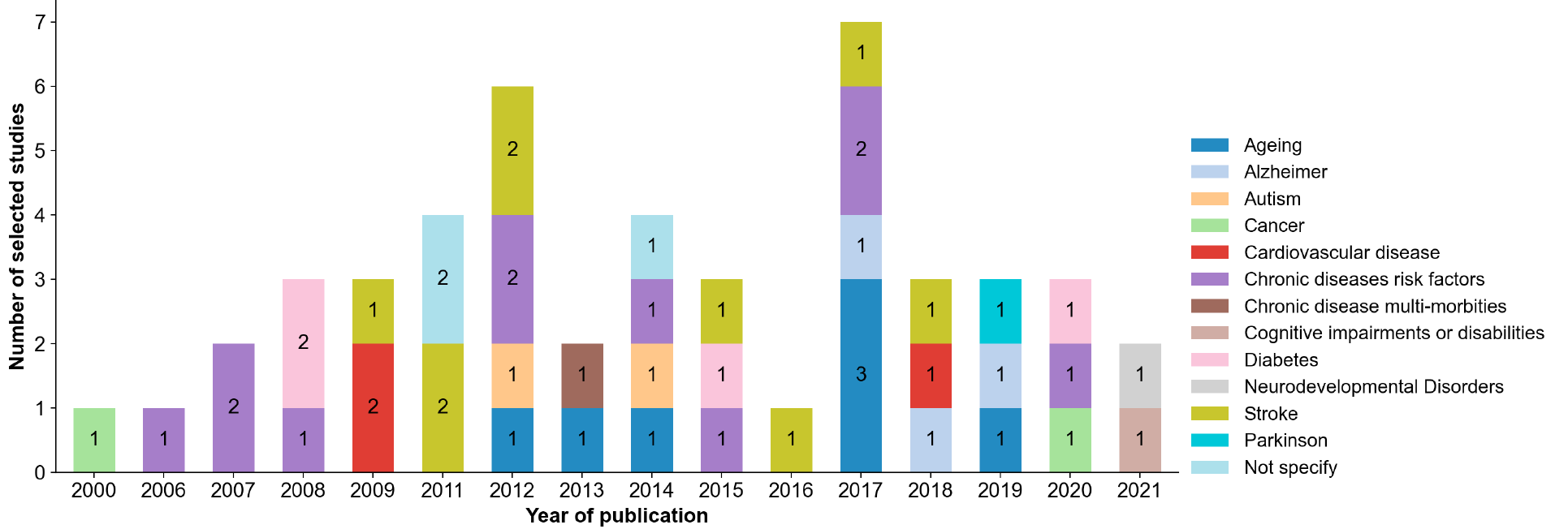}
      \caption{Number of articles selected per year of publication and type of health-related issues.}
      \label{fig:health-relatedyear}
\end{figure}

This RQ focuses on exploring the application of AUIs in the domain of chronic diseases. Of the 48 papers we analysed for this SLR, the majority of the included studies are from European countries (66\%), followed by Canada (11\%). Figure \ref{fig:health-relatedyear} provides a visual representation of the distribution of chronic diseases covered in these 48 primary studies spanning a period of 21 years. The distribution of papers from 2000 to 2021 shows a relatively even distribution, with two minor exceptions. Notably, there is limited research conducted between the years 2000 and 2007. This can be attributed to the early introduction of smartphones in 2006, where the focus was primarily on core ideas and concepts rather than AUI \citep{zhang2022efficacy}.

The use of AUIs in applications for chronic diseases gained traction in 2012, with a substantial increase in the number of related papers. This period also sees an expansion in the range of targeted chronic diseases, encompassing conditions such as stroke, RFCD, ageing, and autism. However, RFCD and stroke have consistently remained the most prominent areas of study over the years. While there appears to have been a decrease in interest in AUI research after 2017, this SLR incorporates recent relevant studies that contribute to the field.
\subsection{Types of software application}  
A key finding in our SLR study, which is also mentioned in other secondary studies \citep{mukhiya2020adaptive, Sanchez2018, Palomares-Pecho2021}, is that most AUI are available on \emph{mobile applications} (18 studies, 57\%) and \emph{web-based applications} (14 studies, 29\%) (see Table \ref{tab:RQ1general}). This can be attributed to the popularity and wide user base of these platforms, making them ideal for AUI implementation. There is also a large number of papers that do not explicitly mention what type of application they proposed (11 studies, 23\%). Only a small number of studies explore \emph{tablet-based applications} (4\%), \emph{desktop applications} (2\%), and \emph{bracelet applications} (2\%), which can be explained by the capability of these platforms. For example, a study that uses a bracelet application as the interface to communicate with the user might be hindered by screen size, thereby limiting the quantity and quality of information presented (S32).

\subsection{Types of solution} Our included studies primarily focus on the development of specific applications, which we categorised based on their primary objectives. Four papers (S14, S27, S34 and S35), for instance, fall under the category of adaptive \emph{algorithms/techniques}, although other studies also include \emph{algorithms/techniques} in the discussion of the application's implementation. A portion of the papers' proposed methodologies or frameworks for generating AUI (15\%) is categorised as \emph{Approach}. One study introduced a tool specifically designed to enhance the adaptability of existing applications (S5). The majority of studies (75\%) proposed different types of applications, with a notable emphasis on \emph{\textbf{health promoting and self-monitoring (HealPM)}} applications (23\%), followed by game applications (17\%). The remaining articles covered \emph{informative, communication, assistive, rehabilitation, and \textbf{healthcare information management (HIM)} applications}, each constituting smaller percentages of the overall distribution.

\subsection{Type of target users} 
\rev{Among the 48 studies, two primary categories of target users are identified, alongside the target users to whom the applications are adapted. It highlights that applications do not universally cater to all app users, contingent upon factors such as user roles and tasks performed through the application. As Table \ref{tab:RQ1general} shows, the first type of target users is the \emph{general public} who have chronic diseases or who are eager to prevent RFCD. The other type is \emph{healthcare professionals} who are responsible for overseeing, monitoring, or managing the health status of other users. 46 of the included studies (81\%) adapted the UI for the general public users, while the remaining 2 studies focus exclusively on adapting it for healthcare professionals. Furthermore, 7 studies (15\%) are designed to target both user groups, with 4 of them solely focusing on adapting for the general public (S2, S13, S15 and S21). This trend can be attributed to the engagement of health professionals in these applications, primarily serving as supervisors and monitors for other users \citep{mukhiya2020adaptive}. As a result, there has been a reduced demand for adaptation catering to health professionals. It is worth noting that the target age groups of applications varied among the studies, with older adults being the most prevalent (13 studies, 27\%), followed by specific age groups such as children, adults, and adolescents in a limited number of studies. However, it is crucial to emphasise that the reporting of demographic information about target audiences in the literature is generally lacking, with over half of the articles (58\%) omitting this information. This observation also suggests that the application is designed to accommodate a diverse range of users, irrespective of their age.}

\section{RQ\textsubscript{2}: How is data being extracted, prepared, and used in the AUI?}\label{sec:RQ2}
In this RQ, we analysed the data types that are modelled and used by various AUIs. The objective of our analysis is to understand the various types of data used, and how the data is collected/extracted. These insights lay the foundation for future research advancements in the effectiveness of using different data sources to develop AUI.

\subsection{RQ\textsubscript{2a}: What types of data are collected to generate an AUI?}\label{sec:RQ2a}
The user, platform and environment that make up the context of use can be seen as key elements in facilitating the adaptive behaviour of the system \citep{Akiki2014, Calvary2003, Norcio1989}. To analyse the types of data being used for generating the AUI, we provided a high-level classification, along with descriptive statistics, outlining the utilisation of specific data types within certain application contexts. Table \ref{tab:dataclass} provides an illustration of different data types used in all included studies, revealing two main types of data sources:  \emph{environmental data} and \emph{user data}. 

{\tiny
     \begin{longtable}{p{10mm}p{11mm}p{4mm}p{12mm}p{13mm}p{45mm}}
     \caption{How AUI are being used.}  
     \label{tab:RQ1general}\\
     \hline
       \textbf{Target users}&\textbf{Adapting users}&\textbf{Study ID}&\textbf{Age group}&\textbf{Type of software application*}&\textbf{Type of solutions} \\
     \hline
    \endfirsthead
    
    \multicolumn{6}{l}%
    {{\bfseries \tablename\ \thetable{} -- continued from previous page}} \\
    \hline
      \textbf{Target users}&\textbf{Adapting users}&\textbf{Study ID}&\textbf{Age group}&\textbf{Type of software application*}&\textbf{Type of solutions} \\
     \hline
    \endhead
    
    \hline \multicolumn{6}{r}{{Continued on next page}} \\ \hline
    \endfoot
    \hline
    \endlastfoot
    \multirow[t]{39}{4em}{General public}&\multirow[t]{39}{4em}{General public}&S1&NS&Mobile&Approach\\
    &&S3&NS&Mobile&Application (Exercise game application)\\
    &&S4&NS&Desktop&Application (Rehabilitation application)\\
    &&S5&Elderly&Web&Tools\\
    &&S6&NS&Mobile& Application (HealPM application)\\
    &&S7&Elderly&Tablet& Application (Therapeutic game application)\\
    &&S9&NS&Mobile& Application (HealPM application)\\
    &&S10&NS&Web&Application (Persuasive game application)\\
    &&S11&NS&Mobile&Application (Rehabilitation application)\\
    &&S14&NS&NS&Adaptive algorithm/technique\\
    &&S16&NS&Web\&Mobile&Application (HealPM application)\\
    &&S17&NS&Web&Application (Informative application)\\
    &&S18&NS&Mobile&Application (HealPM application)\\
    &&S19&Teenager&Mobile&Application (HealPM application)\\
    &&S20&NS&Mobile&Application (HealPM application)\\
    &&S22&Children&Web&Application (Informative application)\\
    &&S23&NS&Mobile&Application (Assistive application)\\
    &&S24&NS&Web&Application (HIM application)\\
    &&S25&Elderly&Web&Application (Assistive application)\\
    &&S26&Elderly&NS&Approach\\
    &&S27&NS&NS&Adaptive algorithm/technique\\
    &&S29&NS&Web&Application (Rehabilitation application)\\
    &&S31&NS&Tablet&Application (Rehabilitation application)\\
    &&S32&Elderly&Bracelet&Application (HIM application)\\
    &&S33&NS&Mobile&Application (Persuasive game application)\\
    &&S34&NS&Mobile&Adaptive algorithm/technique\\
    &&S35&NS&NS&Adaptive algorithm/technique\\
    &&S36&Children&Mobile&Application (HealPM application)\\
    &&S37&Elderly&NS&Application (Exercise game application)\\
    &&S38&NS&Web&Approach\\
    &&S40&NS&Web&Application (Rehabilitation application)\\
    &&S41&Elderly&NS&Application (Assistive application)\\
    &&S42&NS&NS&Approach\\
    &&S43&Elderly&NS&Approach\\
    &&S44&NS&Mobile&Application (Persuasive game application)\\
    &&S45&NS&Web&Application (Persuasive game application)\\
    &&S46&Elderly&Web&Application (Assistive application)\\
    &&S47&Elderly&Mobile&Application (HealPM application)\\
    &&S48&NS&Web&Application (Informative application)\\
    \multirow[t]{2}{4em}{Health professional}&\multirow[t]{2}{4em}{Health professional}&S12&Health professionals&NS&Application (Rehabilitation application)\\
    &&S28&Health professionals&Mobile&Approach\\
    \multirow[t]{7}{6em}{Health professional \& General public}& \multirow[t]{4}{4em}{General public}&S2&Children \& Adult&Mobile&Approach\\
    &&S13&Elderly&Web&Application (HealPM application)\\
    &&S15&Elderly&Mobile&Application (HealPM application)\\
    &&S21&Elderly&Mobile&Application (Communication application)\\
    &\multirow[t]{4}{6em}{Health professional \& General public}&S8&Children&NS&Application (Exercise game application)\\
    &&S30&NS&Web&Application (HIM application)\\
    &&S39&NS&NS&Application (Rehabilitation application)\\
    && & & & \\
    \hline \multicolumn{6}{l}{{* NS: not specified}} \\ 

    \end{longtable}}

\afterpage{
{\footnotesize
    \begin{longtable}{p{17mm}p{25mm}p{65mm}}
    \caption{Classification by type of data source of adaptation}\\
     \hline
     \textbf{Data}&\textbf{Subcategories}&\textbf{Description (studies)}\\
     \hline
    \endfirsthead
    
    \multicolumn{3}{l}%
    {{\bfseries \tablename\ \thetable{} -- continued from previous page}} \\
    \hline
      \textbf{Data}&\textbf{Subcategories}&\textbf{Description (studies)}\\
     \hline
    \endhead
    
    \hline \multicolumn{3}{r}{{Continued on next page}} \\ \hline
    \endfoot
    \hline
    \endlastfoot

    \multirow[t]{6}{8em}{\textbf{User data}\\(User characteristics)}& User's physiological characteristics (29\%)& User's health and normal functioning such as stress level, heart level, blood oxygen level. (S1, S3, S5, S11, S18, S20,  S24,  S27, S29, S32, S39, S41, S42 and S48)\\
    & User's physical characteristics (35\%) & Physical capability and activity level of the user to perform different activities in daily life. (S5, S6, S9, S19, S21, S27, S33, S36, S43, S44, S45, S3, S4, S8, S18, S31 and S39)\\
    & User's psychological characteristics (8\%)& Thoughts, feelings, and other cognitive characteristics that affect a person's mood, attitude, behaviour, and functioning. (S5, S10, S12 and S43)\\
    & User's demographics (15\%)& Quantifiable insights of users into the population such as age, gender, and computer literacy. (S1, S15, S17, S24, S26, S27 and S47)\\
    & User's preference (29\%)& User preferred layout, input/output, theme, interface design. (S13, S16, S22, S23, S26, S30, S46, S17,S19, S21, S24, S25, S28 and S43)\\
    & User's social activity (4\%)& The extent to which the user interacts with others around them. (S9 and S19)\\
    \multirow[t]{4}{8em}{\textbf{User data}\\(Interaction related)}& User's performance in game (31\%)& User's performance in the game includes the scores, success and wins in the game. (S2, S4, S7, S8, S14, S27, S29, S31, S34, S35, S38, S39, S40, S42 and S43)\\
    & User's interaction with the interface (10\%)& History of interactions (eg, click counting, visited, links, time spent, etc). (S14, S25, S13, S37, S46)\\
    & User's emotions (8\%)& User’s emotion when interacting with the interface. (S17, S25, S32 and S37)\\
    & User's feedback (4\%)& User's feedback is collected in the form of questions and answers, or by non-direct methods. (S6 and S11)\\
    \multirow[t]{3}{8em}{\textbf{User data}\\(Task specific)}& User's role (4\%)& User's role is a predefined category assigned to users based on their job title or some other criteria. (S24 and S28)\\
    & User's goals (4\%)& User's goals are the end state(s) that users want to reach. (S12, S29 and S44)\\
    & User's motivation (6\%)& User's motivation is what arouses and sustains action toward a desired goal. (S5, S12 and S34)\\
    \multirow[t]{4}{6em}{\textbf{Environmental data}}& Device types (2\%)& The interface will adapt different devices user used. (S18)\\
    & Operating Platform (2\%)& The interface works across platforms. (S13)\\
    & Environmental condition(s) (4\%)& The interface will adjust e.g. brightness level based on the time of the day and location. (S12 and S28)\\
    & Display sizes (6\%)& The interfaces will adapt to different display sizes providing  the best view to the users. (S12, S13 and S25)
    \label{tab:dataclass}%
\end{longtable}
}}
\subsubsection{User data} User data is categorised into three subcategories as follows: \\

\textbf{User characteristics.} A fundamental requirement for any AUI system is the ability to characterise and differentiate between various end users \citep{Norcio1989}. In our included studies, the most preferred user characteristics used in generating an AUI belong to the  \emph{user's physical characteristic} (35\%),  \emph{user's physiological characteristic} (29\%) and  \emph{user's preference} (29\%), respectively. For example, regarding the physical characteristics, the user’s physical baseline level (weight, height and physical limitations) or actual physical activity level \rev{informs difficulty adjustments for exercise  (S3, S4, S5 and S6) or game (S27 and S8). These data are also harnessed for customising training plans (S21 and S39) or modifying graphic design to motivate sustained physical activity (S19, S33, S36, etc.). Physiological data addressed in the included articles range from heart rate data (S3, S5, S11, etc.), medication treatment (S18), blood glucose level (S20), health impairments (S1 and S18), disease details (S27, S24, S39, etc.) and the blood oxygen level (S42).}

The remaining identified subcategories of user characteristics belong to  \emph{user's demographics} (15\%), \emph{user's psychological characteristics} (8\%) and \emph{user's social activity} (4\%). Notably, cognitive features (S5, S12, and S43) and personality traits (S10) predominantly underlie studies exploring psychological characteristics. In terms of demographic characteristics, age (S1, S24, S27), gender (S27), and literacy (S15, S17, S24, S26, S47) are the most employed data points. Each of these subcategories specifies one dimension of the user characteristics. Our analysis further demonstrates that over 80\% of included articles incorporate at least one of the user characteristic subcategories mentioned earlier, with a notable percentage of studies (29\%) employing multiple subcategories to gain a deeper understanding of users.

\textbf{Interaction related.} An AUI must be able to track user-interface interactions to provide the assistance that is appropriate for the context and the specific user \citep{Norcio1989, Akiki2014, Calvary2003}. Among the identified subcategories of interaction-related data, \emph{user's performance in the game} (31\%) emerges as the most prevalent. The remaining subcategories are divided into  \emph{user's interaction with the interface} (10\%),  \emph{user's emotions} (8\%) and  \emph{user's feedback} (4\%). Our analysis also explores the diversity of interaction-related data dimensions. The findings reveal that approximately half (48\%) of the examined studies incorporate at least one of the above interaction-related subcategories. Moreover, five articles (8\%) leverage data from multiple interaction-related subcategories to comprehensively capture interaction intricacies.

\textbf{Task specific.} \rev{Users can either explicitly define their roles, responsibilities, motivating factors, and goals within the application or allow the system to infer them.} We categorise task-related data into the three subcategories:  \emph{user's roles} (4\%), \emph{goals} (4\%), and  \emph{motivations} (6\%). Among the examined studies, 6 included studies (12\%) use at least one of the above task-specific subcategories.

\subsubsection{Environmental data}
In addition to the data from the user, the adaptive system should learn about itself. This involves knowledge about physical devices (e.g., phone, tablet, laptop, etc.), operating systems and various software applications (e.g., web, desktop, rich Internet application, etc.) \citep{Calvary2003, Akiki2014}. At the same time, the environment in which the device operates contributes to the usage context, and we collectively refer to the above knowledge of the system itself as environmental data. While limited studies mention adaptation to environmental data, certain articles address aspects such as \emph{device type} (S18),  \emph{operating platform} (S13), \emph{environmental conditions} (S12 and S28), and \emph{display size} (S12, S13 and S25).

\subsubsection{Coupling user data}
\rev{Among the primary studies we included, the majority of the studies (58\%) use only one type of user data. Within this group, around 75\% solely incorporate user characteristics data, while the remaining 25\% focus solely on interaction-related data. A substantial portion of studies (42\%) use more than one type of user data to enhance their understanding of users. Among studies employing multiple user data sources, 16 articles (30\%) address user characteristics and interaction-related data concurrently. The most common combination involves \emph{user's physical characteristics} (user characteristics) and \emph{user's performance in the game} (interaction related) data (12\%). Additionally, in 10\% of cases, \emph{user characteristics} are coupled with task-specific data.} The combination of user characteristics and interaction-related data seems to be the most used approach in user profiling. Notably, task-specific data is not used independently across all the examined studies.

\subsubsection{User data and application types}
\begin{figure}[ht!]
  \centering
  \includegraphics[width=\linewidth]{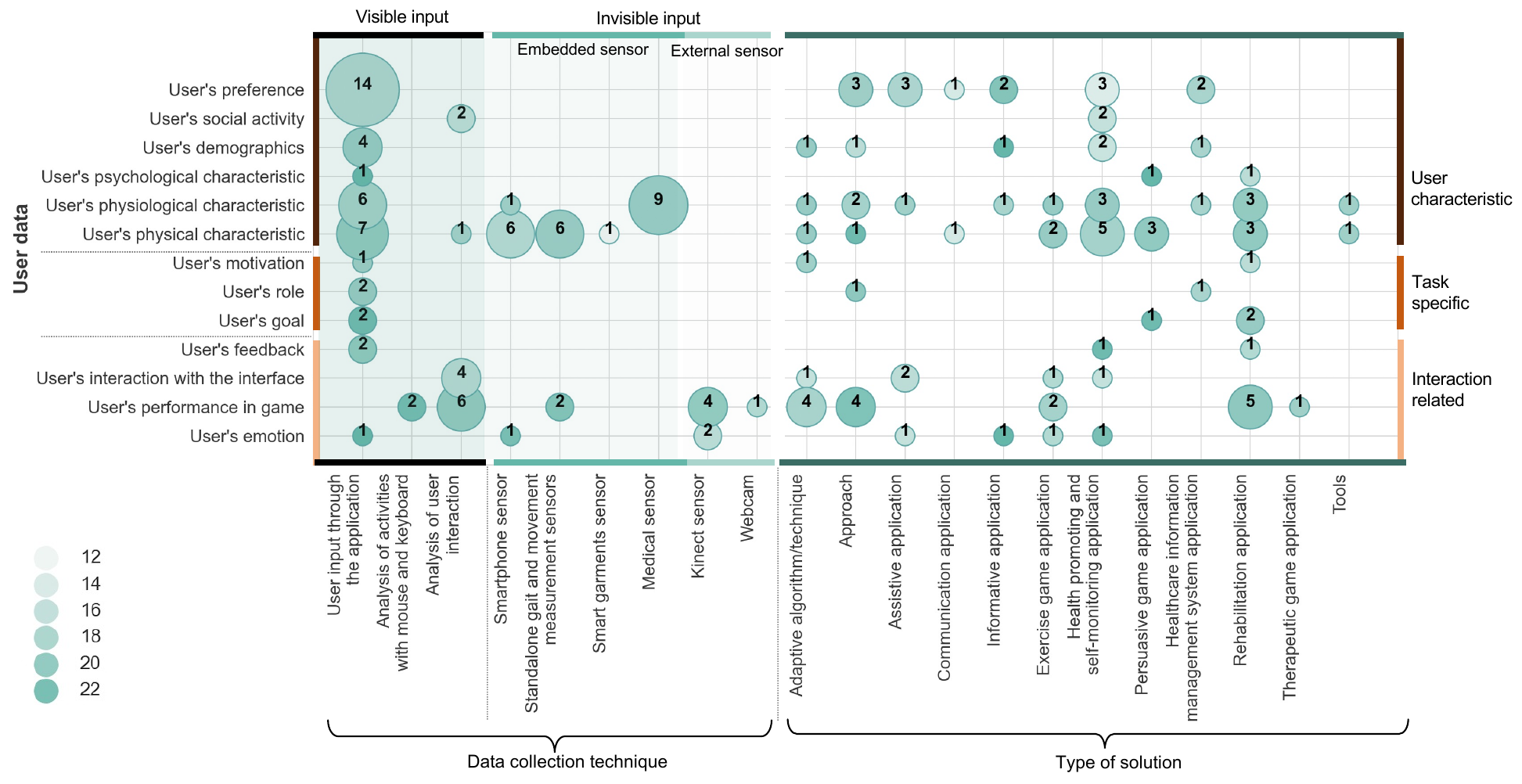}
  \caption{Bubble chart for user data, data collection techniques and type of solution}
%   \Description{Bubble chart for user data and data collection technique}
  \label{fig:datacolbubble}
\end{figure}

 Along with the user data captured, it is critical to comprehend how user data relates to the specific application. Figure \ref{fig:datacolbubble} shows a bubble chart of various types of user data according to the application to which they are applied. The frequency of a certain combination of categories is indicated in each bubble. However, unlike previous review studies \citep{Aranha2021, Palomares-Pecho2021, Sanchez2018}, we added a third dimension to our bubble charts: a legend with a colour scale that indicates the \emph{average study quality score} of all papers in a given bubble. \rev{Examining this figure, we found that the \emph{user characteristics} data is the preferred type of user data used in all kinds of applications, particularly in \emph{rehabilitation and HealPM applications}. Interaction-related and task-specific data have a similar pattern of usage across different applications as user characteristics.}
 
\begin{figure}[hb!]
  \centering
  \includegraphics[width=\linewidth]{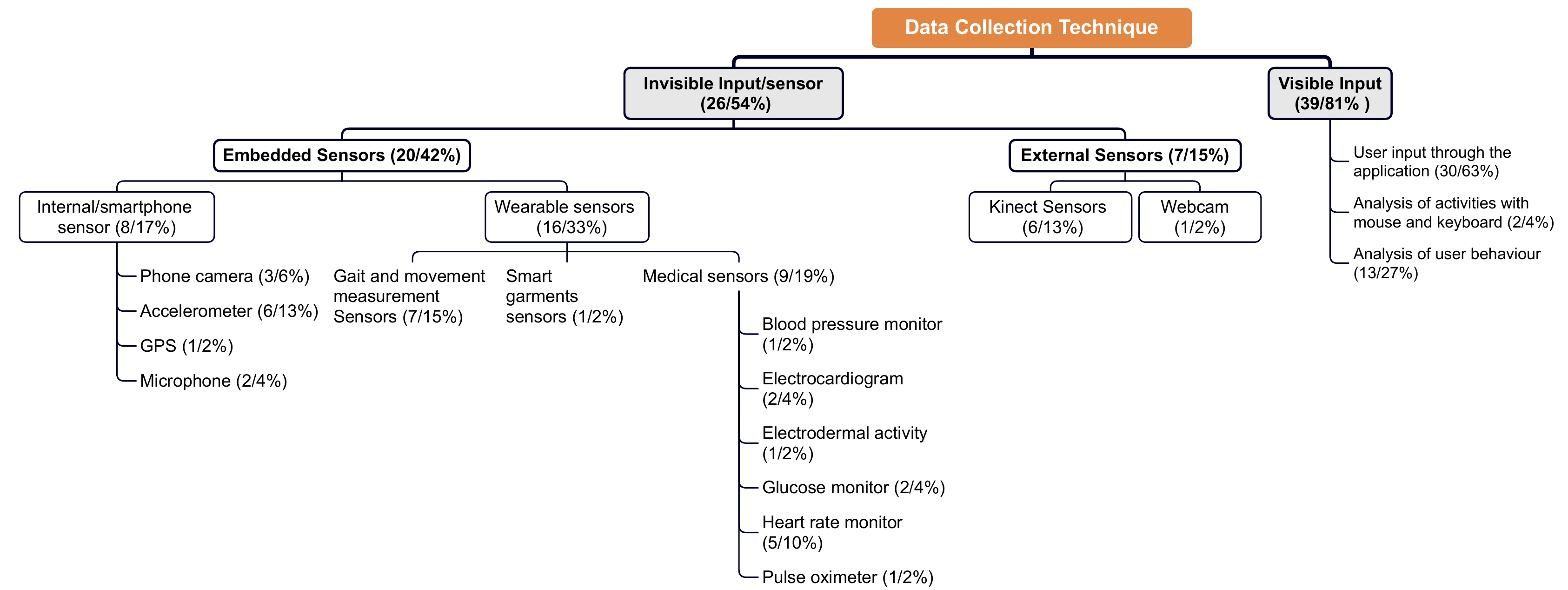}
  \caption{Taxonomy for data collection technique}
%   \Description{Taxonomy for data collection technique}
  \label{fig:datacoltax}
\end{figure}
 
\subsection{RQ\textsubscript{2b}: How is data being extracted/collected?}\label{sec:RQ2b}
In RQ\textsubscript{2b}, \rev{we explored data collection techniques employed to gather and extract user data.  Our investigation into this Sub-RQ leads to the establishment of a taxonomy centred around data collection techniques (See Figure \ref{fig:datacoltax}). Building upon the work of \cite{Aranha2021} and \cite{hurst2012emotion}, we categorised data collection techniques into two primary groups: those based on \textbf{\emph{visible input}} and those based on \textbf{\emph{invisible input}}. Visible input data collection techniques can be observed with the naked eye without the aid of a computer or other technical resources. In turn, invisible input-based data collection techniques mainly involve the analysis of signals and electrical impulses, often requiring specialised sensors for data capture.}

\subsubsection{Data collection techniques based on visible input}
This category involves the acquisition of user data without direct engagement with sensors. In general, visual input is collected by asking the user to answer some questions or by allowing the user to manually change settings and preferences during application usage \citep{hurst2012emotion, Aranha2021}. Among the included studies that use visible input, \emph{user input through application} is by far the main visible input for collecting all types of data, with 30 studies (63\%) using this approach. This approach includes subcategories like questionnaires (S2, S5, etc.), configuring settings manually (S4, S13, etc.), and inputting health/physical data manually (S6, S18, etc.). Another way to use visible input is to use \emph{user behaviour data}, which accounts for 27\% of the included studies. Examples of common forms of behaviour data analysis include examining phone usage patterns (S9, S13, etc.) and assessing user performance in games (S7, S8, etc.). Two studies obtain user information based on activities with a mouse and keyboard (S31 and S38).

\subsubsection{Data collection techniques based on invisible input}
\rev{Apart from the visible input, the invisible input based data collection technique enables seamless data collection without disrupting the user's natural interactions \citep{hurst2012emotion, Aranha2021}. It offers continuous monitoring and alleviates user burden, thereby complementing the capabilities of visible input based data collection techniques. In the identified studies, the invisible input based data collection techniques consist of \emph{embedded sensors} and \emph{external sensors}:}

\textbf{Embedded sensors.} Embedded sensors are usually physically placed in the environment, near the object they monitor, to measure object movement, physiological signals, and environmental variables \citep{Heidemann2004}. The application of embedded sensors is predominantly embedded sensors, with 20 (42\%) of the articles deploying embedded sensors. Among these embedded sensors, \emph{medical sensors} and \emph{smartphone sensors} \rev{stand out as the most frequently utilised, accounting for 19\% and 17\% of the included studies, respectively. Smartphone sensors are employed in various ways, including the phone camera (S20, S32, and S36), microphone (S9 and S20), and accelerometers (S6, S9, etc.).} A diverse range of medical sensors (19\% of studies) are utilised, encompassing heart rate monitors (e.g., S3 and S5), pulse oximeters (S42), glucose monitors (S20 and S41), electrodermal activity sensors (S32), electrocardiograms (S11 and S32), and blood pressure monitors (S29). Additionally, a subset of studies (14\%) employed standalone gait and movement measurement sensors (e.g., S39 and S43).

\textbf{External sensors.} \rev{External sensors are used to capture the user's pose, salient body parts and related objects, and they are less prevalent in all of our included studies compared to embedded sensors \citep{delahoz2014survey}. Among the external sensors, the Kinect sensors (e.g., S8 and S35) and Webcam (e.g., S4) are primarily employed for detecting user motion.}

\subsubsection{Coupling data collection techniques}
 Among the studies that reported their data acquisition techniques (92\%), 20 studies (42\%) used only one data collection technique. Of these, the majority (70\%) employed data collection techniques based on visible inputs. A significant number of studies (38\%) used two types of data collection techniques, and 13\% used three types of data collection techniques. For studies that employed multiple data collection techniques, 8 articles (16\%) collected data through \emph{user behaviour data} and \emph{user input through application}. Furthermore, we found that these two data collection techniques were always used together with invisible input based data collection techniques. This can be explained by the fact that using techniques based on both visible and invisible inputs can identify the needs of the user more accurately and thus improve the quality of adaptation \citep{Aranha2021}. It is also possible that these studies consider users' reluctance to \emph{spend extra time and effort} providing information to the system, as well as the possibility that user-provided information may \emph{not always be accurate} \citep{budzik2000user}. Data collection techniques based on invisible input where the user is not directly involved in the information-gathering process can overcome some of the limitations.

\subsubsection{Data collection technique and user data}
To understand the relationship between different data collection techniques and user data, we computed a bubble chart with bivariate distribution (see Figure \ref{fig:datacolbubble}). Our analysis reveals a strong interdependence between different categories of data collection techniques and types of user data. We observed that user input through applications is by far the dominant method for collecting data on a wide variety of user data. This finding aligns with the expectation that user input serves as the primary technique, constituting 60\% of the studies, for acquiring user data in various contexts. In addition, \emph{user characteristics} such as \emph{user's preferences} (28\% of studies), \emph{physical characteristics} (14\% of studies), and \emph{physiological characteristics} (12\% of studies) are the most popular user data collected through user input. This indicates that researchers predominantly rely on user input to capture and extract \rev{such intricate user characteristic data, which might be challenging to acquire through alternative data collection techniques} \citep{king2013harnessing, pulantara2018development}.

\rev{Figure \ref{fig:datacolbubble} also shows that the external sensors and data collection techniques based on visible input are primarily used to capture the \emph{user's performance in the game}. Another notable trend we observed is that both invisible and visible inputs based data collection techniques can be employed to gather a wide range of interaction related and user characteristic data. For example, user performance data in a game can be collected using the Kinect sensors or by monitoring user interactions. Conversely, task-specific data can only be acquired through techniques based on visible inputs. This discrepancy can be attributed to the inherent difficulty in collecting task-specific data through alternative means \citep{king2013harnessing}.}

\section{RQ\textsubscript{3}: What are the adaptive mechanisms used in generating the AUI?}\label{sec:RQ3}
In this section, we focus on the \textbf{\emph{adaptive mechanism}} covering two key components of the AUI: the \emph{adaptive strategy} and the \emph{adaptation actor}. Our objective is to empirically establish whether specific adaptive strategies are commonly associated with particular application types. Furthermore, we explore the involvement of users during the adaptation process and aimed to develop a taxonomy that encompasses different adaptive strategies, their correlation with adaptation actors, and their association with different applications.

\subsection{RQ\textsubscript{3a}: What types of adaptive strategies are used to generate the AUI?}\label{sec:RQ3a}
Among the studies included in our SLR, we identified a range of adaptive strategies employed for delivering AUI. To organise and categorise these strategies, we utilised the classification proposed by \cite{mukhiya2020adaptive}. Three adaptive strategies are identified:
\begin{itemize}
\item \textbf{Rule-based adaptation:}  The adaptation process in a rule-based approach is characterised by its ability to modify behaviour based on pre-defined rules. This approach offers transparency in monitoring the executed adaptation actions, providing a notable advantage of facilitating the modification of the adaptation process with ease. \citep{mukhiya2020adaptive, jokste2017rule}  (S1, S2, S3, S4, S5, S6, S7, S8, S9, S10, S11, S12, S13, S15, S16, S17, S18, S19, S20, S21, S24, S25, S28, S29, S30, S33, S36, S37, S39, S41, S42, S44, S45, S46, S47 and S48).
\item \textbf{Predictive algorithm-based adaptation:} Predictive algorithm-based adaptation leverages the power of artificial intelligence (AI) techniques to construct sophisticated and objective algorithmic models. These models are specifically designed to analyse user data and generate a comprehensive range of predictions.\citep{mukhiya2020adaptive}. (S9, S14, S19, S27, S31, S32, S34, S35, S37, S38, S39, S40 and S46).
\item \textbf{Adaptation through a feedback loop:} Adaptation through feedback loops is an iterative mechanism employed by systems to dynamically adjust and refine their behaviour based on feedback from the environment or users. This approach involves four essential activities: data collection, analysis, decision-making, and action. Data is gathered from the system and its context, analysed to identify patterns and symptoms, and used to inform decision-making and subsequent actions \citep{krupitzer2015survey, mukhiya2020adaptive, macias2013self, brun2009engineering}. By continuously adapting, the system can optimise its performance over time and align with desired goals or objectives. (S2,S13, S19, S34 and S37).
\end{itemize}

According to our findings, 92\% of the studies report the adaptive strategy they used. 36 studies (75\%) mention that they used rule-based adaptation. The Event-Condition-Action rules, commonly express in the IF-THEN-(ELSE) format, are the most prevalent semantic form of rules in rule-based adaptation systems. For example, in S2, a proposed autism game incorporated visual cues and exciting sensory feedback, such as small firework explosions, to reinforce correct answers. The level of reinforcement is automatically adjusted based on the child's performance. Initially, the system provides high reinforcement at each round of the game. After achieving a streak of 5 correct answers in a row, the reinforcement decreases to every other trial. In the case of errors, the system reverts back to states that offer increased reinforcement. Similarly, in S5, users have the ability to personalise features through the utilisation of trigger-action rules. For instance, if the individual's age exceeds 80, the font size is automatically increased. 

Several studies employed either custom rule languages or adapt existing rule languages to suit their specific needs. For example, in S4, S6, S7, and S8, the activity or performance level thresholds were customised to adjust the difficulty level of the exercise or game. In S6, the physical activity coaching app integrated a rule-based module that aligns training session objectives with established principles of linear progression training \mbox{\citep{american2013acsm}}. This module considers the user's baseline level, progress, perceived fatigue level, and session difficulty level to dynamically adjust the goals and objectives in a personalised manner. In S39, a fuzzy system was employed to dynamically monitor the execution of rehabilitation exercises based on constraints specified by the therapist during configuration. The therapist has the ability to assign a severity level to each constraint, which determines its importance and triggers a corresponding interface adaptation.

 Predictive algorithm-based adaptations are employed in 13 studies (27\%). In S14, an evolutionary algorithm-based optimisation strategy was employed to dynamically adjust the difficulty level of serious rehabilitation games. This approach involves periodic evaluations of the player's perceived difficulty level and subsequent adjustments made within the game to maintain an appropriate level of challenge. In S31, a Monte Carlo tree search algorithm was utilised to perform simulations and identify the best decision among rehabilitation game task sequences, enabling the creation of multiple difficulty modes. Another study (S35) introduced a personalised difficulty adjustment technique utilising reinforcement learning. This technique enables the modification of game properties related to difficulty parameters, such as altering the speed of moving characters within the gaming environment. In S39, the Quest Bayesian adaptive approach was employed to dynamically adjust the gameplay based on the patient's current performance and progress. It is worth noting that certain papers (S27, S38, and S40) did not explicitly specify the algorithms used for adapting the difficulty level of the game. 

 Apart from adapting to the difficulty level of the game, S32 incorporates robust \textbf{\emph{Convolutional Neural Networks (CNNs)}} to accurately detect and recognise the user's emotions. This information is then utilised to generate an adaptive emotional representation within the UI. The adaptive emotional representation is derived from sensor data and the historical evolution of the cognitive assistant platform, which is designed to support active ageing in individuals. In S19, data mining algorithms were utilised to analyse user data obtained from questionnaires and user diaries, enabling the profiling of the user as a gamer and enhancing their overall enjoyment and satisfaction with the health-promoting system designed for teenagers. In S37, decision trees were utilised to classify users based on the values of various attributes, including their interactions with the system and heart rate. By categorising users into different classes, the study created a set of preferred interface configurations tailored to each user type.
 
 A small proportion of studies use adaptation through a feedback loop approach, representing 7\% (3 studies) of the included studies. S2 introduced a method for adapting study quiz prompts to align with specific teaching concepts. The system employed continuous monitoring of the user's quiz performance and dynamically generated diverse prompts tailored to the individual's needs. In the case of an assistive application designed for the elderly (S13), the system incorporated a tracking mechanism to identify the user's most frequently accessed elements over a period of two weeks. Based on this information, the system dynamically rearranges the elements within the interface. This iterative process of monitoring and adaptation ensures that the interface remains responsive to the user's preferences and evolving needs. In S34, a rehabilitation game utilised more than just algorithmic approaches to adapt the difficulty levels. These games incorporate a feedback mechanism that gathers user feedback through a logistic-type Likert scale. This scale enables users to express their impressions of the previous game session, including factors such as boredom, motivation, and stress conditions. By incorporating this feedback, the game aims to enhance its adaptive strategies and optimise the overall user experience.
 
Of the studies that reported adaptive strategies, the majority (71\%) employ a single strategy to accomplish the required adaptation. Among the single-strategy studies, rule-based adaptation (79\%) is the most prevalent adaptation method, while the remaining 20\% of the studies used adaptive methods based on prediction algorithms. The popularity of rule-based adaptation can be attributed to its relative simplicity compared to other complex adaptation strategies \citep{he2006flexible}.

\subsubsection{Coupling adaptive strategy}
For the studies that employ multiple adaptive strategies, 6 articles (12\%) using rule-based adaptation also use predictive algorithm-based adaptation. In 7\% of the included studies, rule-based adaptation is accompanied by adaptation through a feedback loop. Therefore, it appears that the combination of rule-based adaptation with other adaptive strategies is predominantly employed in studies involving AUI. The utilisation of a feedback loop approach for adaptation is consistently accompanied by other strategies, indicating that this approach is rarely employed in isolation.

\subsubsection{Adaptive strategies and applications}
\begin{figure}[hb!]
    \centering
     \includegraphics[width=0.85\linewidth]{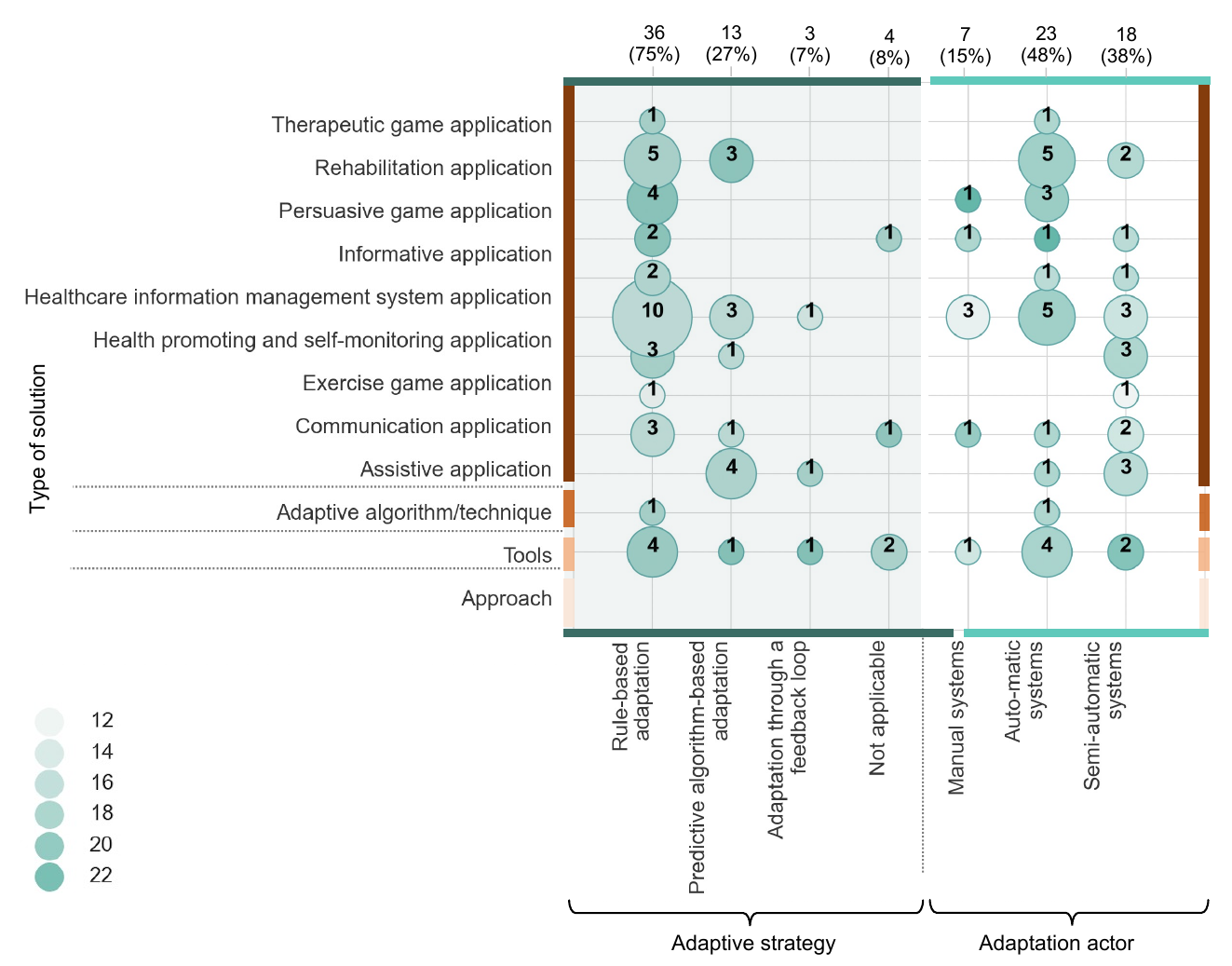}
    %   \Description{Co-occurrence matrix between categories of adaptive strategy}
  \caption{Bubble chart for adaptive mechanism and types of solution.}
  \label{fig:typeandstrategy}
\end{figure}

Figure \ref{fig:typeandstrategy} delineates a bivariate distribution of various types of adaptive strategies according to the application to which they are applied. Examining this data, we found that rule-based adaptation is by far the leading adaptive strategy for various kinds of applications. We also observed that adaptation through a feedback loop and predictive algorithm-based adaptation have mostly been used in \emph{rehabilitation applications, HealPM applications and exercise game applications}. One possible reason for this observation is that these applications require the ability to dynamically adjust the difficulty level of the game or rehabilitation exercises in order to sustain the user's motivation (S3, S6, S7, and S14). At the same time, it is crucial for the exercise game to ensure that the actual outcome of the workout aligns with the user's expectations regarding their motor skills \citep{munoz2019kinematically, skjaeret2016exercise}. Consequently, the application may prompt the user for feedback and make adaptations accordingly to maintain a desirable user experience.

\subsection{RQ\textsubscript{3b}: What roles do different actors play in AUI adaptation?}\label{sec:RQ3b}
For RQ\textsubscript{3b}, our primary focus is exploring the involvement of the user in the adaptation process. To investigate this, we have categorised the adaptation into three categories \emph{(manual systems, automatic systems and semi-automatic systems)} based on whether the system or the end-user takes the initiative for adaptation, as discussed in Section \ref{subsec:adaptivuser}. Our  analysis shows that nearly half (48\%) of the papers employed the \emph{automatic system} approach to construct the AUI (S2, S5, S7, S9, S11, S12, S17, S18, S20, S24, S26, S29, S31, S32, S33, S35, S36, S38, S40, S41, S42, S44 and S45). On the contrary, few studies(15\%) reports the use of \emph{manual system} approach (S1, S10, S15, S16, S22, S23 and S47). Notably, the \emph{semi-automatic systems} approach features in 18 studies, constituting 38\% of all included studies (S3, S4, S6, S8, S13, S14, S19, S21, S25, S27, S28, S30, S34, S37, S39, S43, S46 and S48). 

\subsubsection{Adaptation actor and applications} We observed certain trends between the adaptation actors and their associated applications, as depicted in Figure  \ref{fig:typeandstrategy}. A prevalent pattern is the utilisation of \emph{semi-automatic system and automatic system} across a spectrum of applications. This pattern aligns with expectations, as manual systems place the entire responsibility of adaptation solely on the user \citep{Norcio1989}. Moreover, empirical evidence suggests that users often struggle to effectively navigate the adaptable features of manual systems, leading to infrequent usage \citep{Norcio1989, Oppermann1994}. We also observed an imbalanced distribution of studies utilising automatic systems, with a higher proportion observed in \emph{rehabilitation applications, persuasive game applications, and HealPM applications}. In these types of applications, user engagement and active participation play a significant role in facilitating desired outcomes, such as improving rehabilitation progress, motivating behaviour change, promoting health, and enabling self-monitoring, automatic systems are advantageous. These systems seamlessly adapt to user needs and preferences based on collected interaction data, thereby bolstering user participation and maximising the efficacy of these applications \citep{Harman2014, Oreizy1999}. As a result, the utilisation of automatic systems becomes particularly advantageous in application domains where user engagement is a key requirement for successful achievement.

\subsubsection{Adaptation actors and adaptive strategy}
We also analysed the relationships between adaptive strategy and different adaptation actors within a given application. \rev{Applications that adopt adaptive strategies, such as predictive algorithm-based adaptation or adaptation through a feedback loop, tend to align with the realm of \emph{automatic or semi-automatic systems}. This stems from the inherent attributes of these strategies that require \emph{complex user input} and \emph{regular UI adaptation}. Such processes are more efficiently automated rather than being reliant on user operation. For example, tasks such as adjusting the difficulty level and maintaining user performance at an acceptable level demand constant extraction, processing, or calculation of data, better handled by automated systems.}

Additionally, users often lack the necessary knowledge and expertise to modify the interface and may be unaware of the possibility or potential benefits of making changes themselves \citep{Norcio1989}. Another correlation we observed is the diverse nature of interface adaptation through rule-based adaptation. In this approach, users have the flexibility to manually adjust the parameters for different variables, prompting corresponding changes in the interface. Conversely, certain solutions employ dynamic interface changes automatically driven by system-defined rules \citep{he2006flexible}. 

\section{RQ\textsubscript{4}: What are the adaptive elements used in the AUI?}\label{sec:RQ4}
The objective of this RQ is to explore the current utilisation of adaptive elements and examine their usage in various health-related applications. We categorise the adaptive elements into three main groups: \emph{presentation adaptation, content adaptation, and behaviour adaptation} (as shown in Table \mbox{\ref{tab:adaptationclass}}). The subcategories within these main adaptive elements, discussed in this section, represent the most prevalent findings among the included articles, which may not necessarily encompass all possible classifications in the field.

\afterpage{
{\footnotesize
    \begin{longtable}{p{15mm}p{25mm}p{65mm}}
    \caption{Classification by adaptive elements}\\
    \hline
    \textbf{Type of adaptation}&\textbf{Subcategories}&\textbf{Description (studies)}\\
     \hline
    \endfirsthead
    
    \multicolumn{3}{l}%
    {{\bfseries \tablename\ \thetable{} -- continued from previous page}} \\
    \hline
    \textbf{Type of adaptation}&\textbf{Subcategories}&\textbf{Description (studies)}\\
     \hline
    \endhead

    \hline \multicolumn{3}{r}{{Continued on next page}} \\ \hline
    \endfoot
    \hline
    \endlastfoot
    
    \multirow[t]{3}{8em}{Presentation adaptation}&Graphic design (29, 60\%)& Change the layout, font size, colour, and theme. (S1, S4, S5, S7, S8, S9, S10, S12, S13, S14, S15, S16, S18, S20, S21, S22, S23, S24, S25, S28, S30, S32, S33, S36, S39, S42,S43, S44, S45, S46 and S47)\\
    & Information architecture (2, 4\%)& Change the structural design of information. (S18)\\
    & Sound effect (1, 2\%)& Change the volume of the sound. (S13)\\
    \multirow[t]{2}{6em}{Content adaptation}&Interface Elements rearrangement (2, 4\%)& Change the interface elements by removing, adding, or rearranging. (S13 and S48)\\
    & Content complexity (5, 10\%)& Change the content complexity so that it is easy for individuals to understand and process based on users' cognitive skills, educational backgrounds, comprehension capabilities and other qualities. (S12, S15, S17, S41 and S47)\\
    \multirow[t]{5}{8em}{Behaviour adaptation} & Navigation adaptation (2, 4\%)& Change the user's permission to navigate freely or navigation to other modules is suppressed. (S20 and S25)\\
    & Add on functions (6, 13\%)& Add new functions to better assist user use the application, e.g., magnifying function for ageing users (S1, S16, S20, S24, S25 and S30)\\
    & Different persuasive strategy (4, 8\%)& Change pervasive strategies used to motivate the desired behaviour change, according to different user type or status. (S5, S10, S19, S37 and S41)\\
    & Difficulty level (23, 48\%)& Change the difficulty level of the game or exercise based on the motivation state or user performance. (S2, S3, S4, S6, S7, S8, S10, S11, S14, S19, S26, S27, S29, S31, S34, S35, S37, S38, S39, S40, S41, S42 and S43)\\
    & Multimodal interaction (6, 13\%)& Change the modality of the interface based on different contexts of use. (S1, S20, S25, S28, S37 and S43)
    \label{tab:adaptationclass}
    \end{longtable}}}

\subsection{Types of Adaptation}
\textbf{Behaviour adaptation.} \rev{Our analysis identifies that the most prevalent adaptive element is behaviour adaptation. This involves modifying navigation type or structure, activation or deactivation of interface elements, and adjustments to interaction modalities within the application} \citep{Paterno2012,  Potseluiko2021, Vasilyeva2005}. We categorise behaviour adaptation as shown in Table \ref{tab:adaptationclass}. Among the subcategories, \emph{difficulty level} (48\%) is by far the leading adaptive element within behaviour adaptation. \rev{This finds significant use in games, exercise, and rehabilitation applications, aiming to sustain user engagement by tailoring challenges to their physical capacity and health status} \citep{Aranha2021}. Changes in difficulty level take into account elements such as changing the number or size of puzzle pieces in the game (S7) and target speed during the exercise (S14, S35 and S39). Additionally, two other common adaptive elements within behaviour adaptation are \emph{multimodal interface} adaptation (13\%) and \emph{add-on functions} (13\%). In S28, when the physician visits the patient's room, the applications switch to employing voice input. This adaptation enables the physician to efficiently search for the patient's medical record number or order a lab test. In S30, the add-on module "Magnifying Glass" can be triggered after processing the user's health information. 

The remaining infrequent adaptive elements include \emph{different persuasion strategies} (8\%) and \emph{navigation adaptation} (4\%). In S10, the authors aimed to tailor the persuasive strategies of a healthy eating game to different player types. They developed two versions of the game: one utilising the \emph{reward} strategy and the other employing the \emph{competition} strategy. In the version using the reward strategy, players are shown their current point score and received badges as a form of acknowledgement for their achievements within the game. On the other hand, the competition-based version introduces a simulated leaderboard that compares players' performance and displays their names, scores, and ranks. In S20, a mHealth app was developed for the management of hypoglycemia. The application features a standard homepage that provided patients with the freedom to navigate through various visualisations, including connected sensor data, medical test results, and exercise results. However, when there is a possibility of a hypoglycemic event occurring, the navigation to other modules within the application is intentionally suppressed. Instead, the application emphasises quick access to emergency services by highlighting the dedicated feature, ensuring that patients could swiftly access the necessary assistance during a hypoglycemic episode.

\textbf{Presentation adaptation.} Presentation adaptation refers to changing the parameters of the interface elements such as colour, size, the position of objects, font size, etc.\citep{Paterno2012,Potseluiko2021,Vasilyeva2005}. We break presentation adaptation down as shown in Table \ref{tab:adaptationclass}. \emph{Graphic design} adaptation is performed in 63\% of all included studies. This approach can be applied in many ways, such as controlling the theme, layout, and display of the application. Other possibilities considered in the analysed articles are the change of \emph{information architecture} or \emph{sound effects} which account for 2\% and 2\% of all included studies, respectively. \emph{Information architecture} refers to the variation in the structural design of information. For example, in S18, a mHealth system was designed to cater to the individual needs of patients with both diabetes and cardiovascular disease. The system aims to personalise the self-care process by providing tailored messages to patients, which is achieved by adjusting the nature, form, and structure of the information presented to patients, taking into account their profile, medical treatment goals, and specific content. \emph{Sound effect} is an adaptive element in a single included study (S13). In that study, the output volume of sound effects is adjusted based on the distance between the user and the device.

\textbf{Content adaptation.} Content adaptation involves modifying the content level of the interface by adjusting the text, its semantic content, images or explanatory inscriptions \citep{Aranha2021,  Palomares-Pecho2021, Paterno2012, Potseluiko2021, Vasilyeva2005}. The adaptation of interface content is less common among the primary studies that we examined. Specifically, we found that the adaptation of \emph{content complexity} and \emph{interface elements rearrangement} accounted for 10\% and 4\% of all included studies, respectively. Adaptation at the level of \emph{content complexity} seeks to make content simple to grasp based on users' cognitive skills, educational backgrounds or comprehension capabilities. For instance, in study S15, a personal health monitoring application is developed to address various health-related issues, including sleep problems, diet, and pain management. The application's interface is designed to accommodate users with different skill levels, allowing for customisation to meet the needs of older individuals with limited IT knowledge. As part of this adaptation, the interface features a single-button layout to enhance accessibility for older users unfamiliar with the technology. Additionally, two studies, S15 and S48, discussed the rearrangement of interface elements, wherein the most frequently accessed items were positioned at the top of the corresponding menus. 

\subsection{Combining multiple adaptive elements}
We identified 30 (63\%) studies that utilised only one type of adaptive element in their interface design.18 studies (38\%) incorporate multiple types of adaptive elements. Among the studies that employ multiple adaptive elements, 13 (27\%) utilise both \emph{presentation adaptation} and \emph{behaviour adaptation}. When examining the subcategories of adaptive elements, we observed that \emph{graphic design} is the most commonly utilised subcategory in combination with other adaptive elements within presentation adaptation. Conversely, other subcategories of presentation adaptation, such as \emph{information architecture} and \emph{sound effects}, are less frequently used in conjunction with other types of adaptive elements. In contrast, subcategories under behaviour adaptation are commonly employed together. For instance, \emph{multimodal interaction} adaptation is often combined with \emph{navigation, add-on functions, different persuasive strategies, and difficulty levels}, accounting for 4\%, 6\%, 2\%, and 4\% respectively.

\subsection{Adaptive elements and chronic diseases}
\begin{figure}[!h]
\centering
      \includegraphics[width=0.85\linewidth]{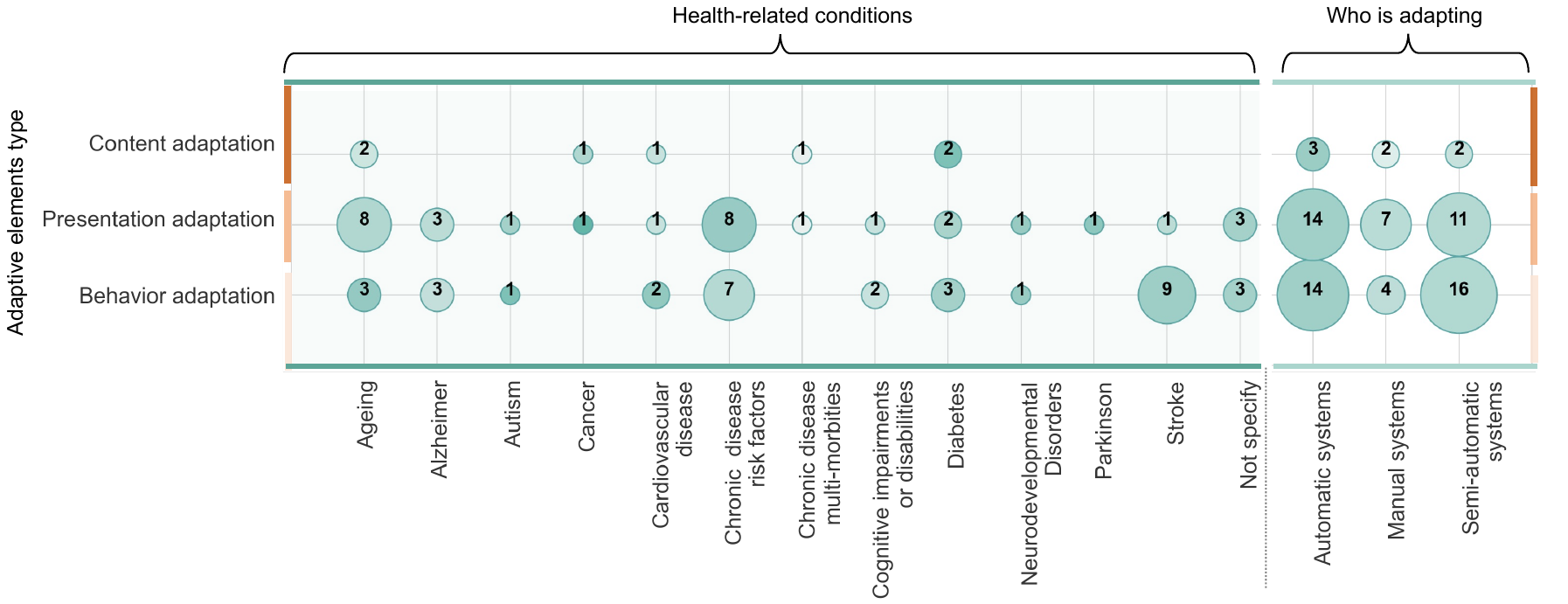}
      \caption{Bubble chart for health-related conditions, adaptation types and adaptive elements.}
     \label{fig:bubbleadaptelem}
\end{figure}

 Our analysis indicates that \emph{adaptation of presentation and behaviour on the interface} are the two main types of adaptation used in studies supporting the \emph{stroke, RFCD and ageing} (see Figure \ref{fig:bubbleadaptelem}). These health-related conditions are often addressed through applications designed for rehabilitation, health monitoring, and exercise. Notably, behaviour adaptation is implemented in these applications to enhance user engagement and interaction. Moreover, it is observed that graphic design is frequently incorporated alongside various subcategories of behaviour adaptation. This observation suggests a potential explanation for the prevalence of presentation and behaviour adaptation in the context of stroke, RFCD, and ageing.

\section{RQ\textsubscript{5}: How is the AUI developed and evaluated?}\label{sec:RQ5}
In this RQ, our focus is on investigating the approaches employed in the design and evaluation of chronic disease-related applications with AUI. The use of scientific and rigorous evaluation methods to evaluate software development methods has long been emphasised by the SE community \citep{Hachey2012, Zannier2006}. In our study, we explore the evaluation of AUI through three critical lenses: \emph{evaluation approach, evaluation measurements and evaluation results}. 

\subsection{Design approach}
Although the design methodology for AUIs plays a crucial role in their development, the included literature is disappointingly scant on this matter. According to our findings, only 69\% of articles included in our SLR reported the specific design approaches they employed to create adaptable interfaces. Of these, 21\% followed a \emph{model-driven approach} \citep{soley2000omg}, and only 3\% were based on a \emph{user-centred design approach} \citep{abras2004user}. An \emph{iterative design approach} \citep{iterative} and \emph{inclusive design approach} \citep{coleman1999inclusive} were used by 2\% each. In conventional programming practices, developers typically construct software by manually composing lines of code within a programming language. In contrast, the model-driven approach involves the utilisation of graphical or textual models that depict the intended behaviour and structure of the system. Subsequently, these models undergo automated transformations to generate executable code \citep{soley2000omg}. Given the inherent benefits a model-driven approach provides, such as enhanced traceability, technology independence, and the ability to perform comprehensive outcome evaluations for the AUI \mbox{\citep{Akiki2014, Vasilyeva2005}}, we expect that the utilisation of model-driven approaches in the development of AUIs will gain significant popularity in the academic community and beyond.

In our investigation of studies employing a model-driven approach, we examined the various types of models utilised in the design of AUI. Our analysis reveals a spectrum of models, including the context model (S5), user model (S19, S36, and S42), ontology model (S1), personalisation model (S10), player model (S14), adaptation model (S20), and user behaviour model (S14). Notably, the context model and user model emerged as the most frequently referenced models across different solutions.

\subsection{Evaluation approach} We adopted the \emph{evaluation approach classification scheme} introduced by \cite{Chen2011} to categorise the evaluation approach employed in the included studies. Figure \ref{fig:bubbleevalproto} presents the distribution of evaluation approaches used in the included papers. \emph{“Experiment with human subjects”} (56\%) is the most frequently used means of evaluation. This is followed by \emph{"example application"} and \emph{"discussion"} with 15\% and 10\%, respectively. \emph{"Field experiment"} is only applied to 8\% of the total evaluated papers. Other less common methods include \emph{"simulation"} (4\%), \emph{"experience"} (2\%), and \emph{"experimenting with software"} (4\%). For the evaluation method involving human participants, we found that the majority of studies (33\%) have a participant count ranging  \emph{from 1 and 10} (see Table \ref{tab:evaluationdet}). The "11 to 20" and "21 to 30" participant groups account for 10\% each, while only a minority of studies include more than 31 participants (8\%). It is worth noting that none of the included studies provides explicit justification for their chosen number of participants, despite the extensive research within the HCI field on determining the ideal number of participants for different study types \citep{Hachey2012, Sanchez2018}. 

In general, the evaluation of systems with AUI necessitates a \emph{robust and rigorous methodology}. Rigorous evaluations not only enhance the credibility of the study but also contribute to the advancement of AUI research by generating robust and replicable findings. Upon further examination, we note that studies that employ human subjects for evaluation consistently achieve high study quality scores, particularly those conducted in real-world environments such as field settings, homes, or hospitals (see Figure \ref{fig:bubbleevalproto}). By involving human subjects in real-world contexts during the evaluation process, researchers can observe user interactions with the AUI system in natural and authentic settings, capturing a broader range of scenarios and uncovering potential challenges or opportunities for improvement.

\subsection{Evaluation Measurements}
\begin{figure}[ht!]
  \centering
  \includegraphics[width=\linewidth]{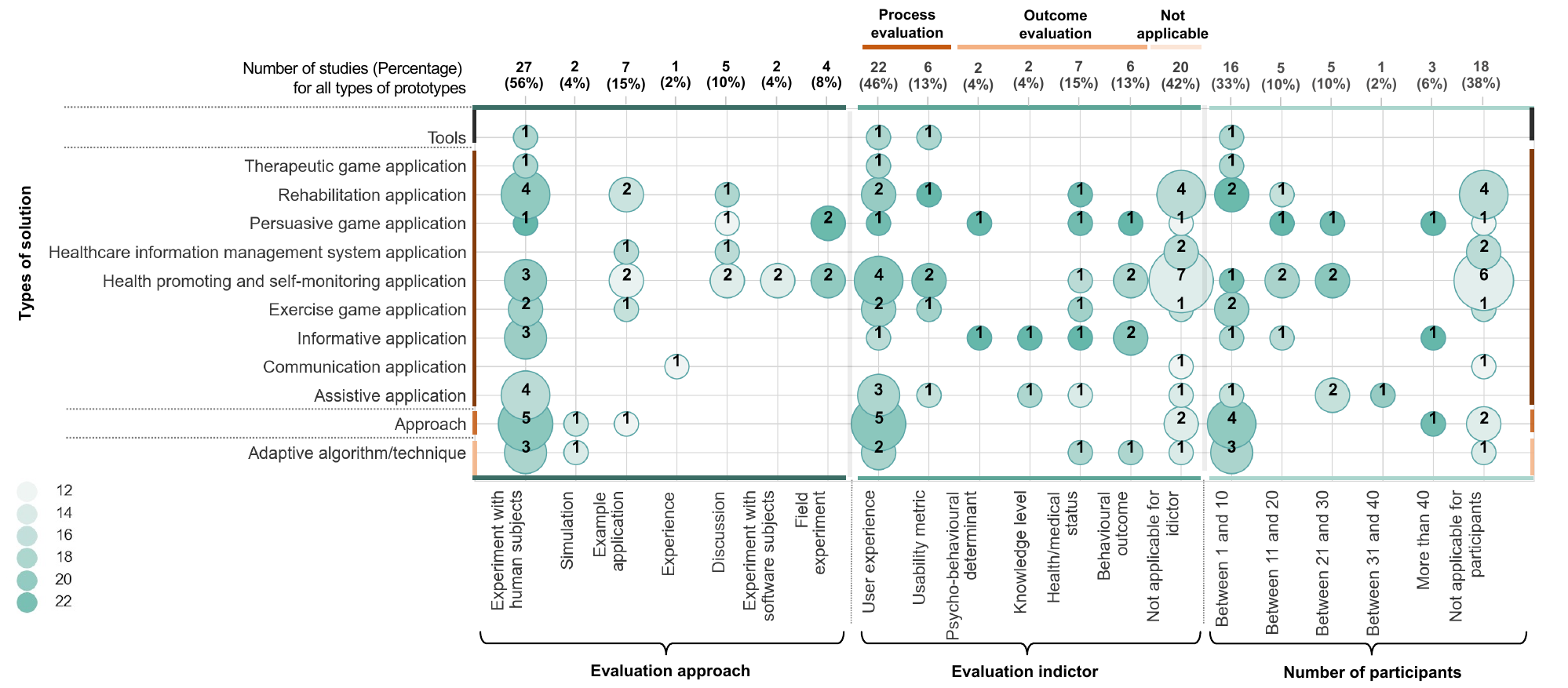}
  \caption{Bubble chart for the type of solution, evaluation approach, evaluation indicator and number of participants}
  \label{fig:bubbleevalproto}
\end{figure}

We focus on evaluation methods that provide substantial evidence and excluded studies that rely on weak evaluation methods such as "example applications", "discussion", and "experience" \citep{Chen2011}. As a result, we only considered 31 studies (62\%) for this analysis. Based on our analysis and referring to the classification made in \cite{KamelGhalibaf2019}, we categorise the evaluation indicators into six main categories: \emph{behavioural outcome, health/medical status, user experience, psycho-behavioural determinant, knowledge level, and usability metric} (see Table \ref{tab:evaluationdet}). According to the Centers for Disease Control and Prevention (CDC) \citep{CDC}, these six categories can be summarised into two general categories, mainly \textbf{\emph{process evaluation}}  and \textbf{\emph{outcome evaluation}}. Outcome evaluations (including behavioural outcome, health/medical status, psycho-behavioural determinant, and knowledge level) are conducted to appraise the efficacy of a program in generating change. Conversely, process evaluations (including user experience and usability metrics) provide insights into the process through which program outcomes or impacts were attained. Following is a breakdown of included studies by evaluation indicator categories: 22 studies (46\%) employ process evaluation while 13 studies (27\%) use outcome evaluation. Seven (15\%) of the articles undergo both process and outcome evaluation, aiming to leverage early feedback from the process evaluation to improve the design of the outcome evaluation. To identify the most common evaluation metrics, a thorough review of these methods is conducted as follows. Six studies (13\%) of the 22 (46\%) studies employ process assessment evaluated based on usability metrics. User experience is assessed in all included studies using process evaluation. Of the 42\% of included studies evaluating outcomes, 19\% focus on behavioural outcomes, 16\% examine health/medical status, 6\% assess psycho-behavioural determinants and 6\% measure the knowledge level of users. Figure \ref{fig:bubbleevalproto} indicates a preference for process evaluation (50\%) across various types of applications.

\subsection{Evaluation Results}
 Among the included studies, we found that some studies do not provide clear and specific statements about the evaluation result. Many primary studies discuss the advantages and disadvantages of the solution without explicitly reporting the effects of the solution. Some studies use real examples and collect evidence of its use informally, only describing how to use a particular method without explaining any specific effects of using the method. In general, the evaluation results show that in 92\% of articles tailoring was \emph{effective}, in 4\% \emph{not specify} and in 4\% it was \emph{partially effective}.

{\tiny
 \begin{longtable}{p{13mm}p{5mm}p{12mm}p{25mm}p{10mm}p{10mm}p{14mm}}
 \caption{Evaluation details of studies involve human subjects.}  \label{tab:evaluationdet}\\
 \hline
    \multirow[t]{2}{10em}{\textbf{Health relate conditions}}
     &\multirow[t]{2}{3em}{\textbf{Study ID}}
     &\multirow[t]{2}{8em}{\textbf{Evaluation indicator *}}
     &\multirow[t]{2}{12em}{\textbf{Length of time evaluation}} 
     &\multirow[t]{2}{4em}{\textbf{Evaluation results}}
     &\multicolumn{2}{p{24mm}}{\textbf{Users talking part in studies}} \\ \cline{6-7}
     & & & & & \textbf{General users} & \textbf{Health professionals} \\
     \hline
\endfirsthead

\multicolumn{7}{l}%
{{\bfseries \tablename\ \thetable{} -- continued from previous page}} \\\hline
    \multirow[t]{2}{8em}{\textbf{Health relate conditions}}
     &\multirow[t]{2}{3em}{\textbf{Study ID}}
     &\multirow[t]{2}{8em}{\textbf{Evaluation indicator}}
     &\multirow[t]{2}{12em}{\textbf{Length of time evaluation}} 
     &\multirow[t]{2}{4em}{\textbf{Evaluation results}}
     &\multicolumn{2}{p{24mm}}{\textbf{Users talking part in studies}} \\ \cline{6-7}
     & & & & & \textbf{General users} & \textbf{Health professionals} \\
     \hline
    \endhead
    \hline
    \multicolumn{7}{r}{{Continued on next page}} \\ \hline
    \endfoot
    
    \hline
    \endlastfoot

    \multirow[t]{6}{3em}{Stroke}&S27&UX&1 session per participant & Effective  &8 adults&  \\
    &S31&UX\&UM&2 week (one session per day) per participant&Effective&7 adults&  \\
    &S34&BehOut&10 sessions (1 minute each) per participant& Effective &5 adults&  \\
    &S35&UX\&HeathSta&8 session ( 20 minutes) per participant&Effective&8 adults&  \\
    &S38&UX&2 session (2 minutes) per participant &Effective &10 adults& \\
    Parkinson disease&S23&UX&1 session per participant (30 minutes)&Effective&40 adults&  \\
    Not specify&S28&UX&1 session per participant (1 day)&Effective& &45 docotors\\
    Neurodevelop- mental disorders&S8&UX \&UM&1 session per participant & Effective &9 adults & 6 therpists\\
    \multirow[t]{3}{2em}{Diabetes}&S17&HeathSta, BehOut, PsyDet\&KnowL&1 session per participant & Partially effective. &561 adults&  \\
    &S18&HeathSta\& BehOut&1 session per participant &Not specify&15 adults& \\ 
    &S41&UX\&KnowL&1 session per participant &Effective&28 adults&  \\
    \multirow[t]{7}{4em}{RFCD}&S3&UX\& HeathSta&1 session per participant (10 minutes)&Effective&2 children&  \\
    &S9&BehOut\&UX&19 days& Effective &27 adults&  \\
    &S10&PsyDet&1 session per participant (20 minutes)&Effective&272 adults&  \\
    &S36&UX\&UM&1 session per participant &Effective&12 children&  \\
    &S42&UX&1 session per participant &Effective&8 adults&  \\
    &S44&HeathSta\& UX&3 months &Effective&28 adults&  \\
    &S45&BehOut&14 weeks&Effective&19 adults&  \\
    Cardiovascular disease&S29&HeathSta&43 sessions (at least 30 minutes each) per participant&Effective&10 adults&  \\
    \multirow[t]{2}{3em}{Cancer}&S6&UX \&UM&2 sessions per participant &Effective&22 adults&  \\
    &S48&UX&Not specify& Partially effective&18 adults&  \\
    \multirow[t]{2}{3em}{Autism}&S2&UX&1 session per children participants (30-60 minutes)&Effective&7 children&2 therapists\\
    &S22&BehOut&15 weeks&Effective&9 children&  \\
    \multirow[t]{3}{3em}{Alzheimer}&S7&UX&1 session per participant &Effective&10 adults &  \\
    &S25&HeathSta& 8 sessions &Effective& 25 adults&  \\
    &S46&UX\&UM&1 session per participant &Effective&8 adults&  \\
    \multirow[t]{3}{3em}{Aging}&S5&UX\&UM&2 sessions per participant & Effective&3 adults&4 caregivers\\
    &S32&UX&Not specify&Effective&  &10 caregivers\\
    &S43&UX&Not specify&Effective&63 adults&  \\
    \hline \multicolumn{7}{p{100mm}}{* behavioural outcome (BehOut), health/med-
ical status (HeathSta), user experience (UX), psycho-behavioural determinant
(PsyDet), knowledge level (KnowL), and usability metric (UM). } \\ 
    \end{longtable}
}
\section{Discussion}\label{sec:discussion}
\begin{figure}[ht!]
  \centering
  \includegraphics[width=\linewidth]{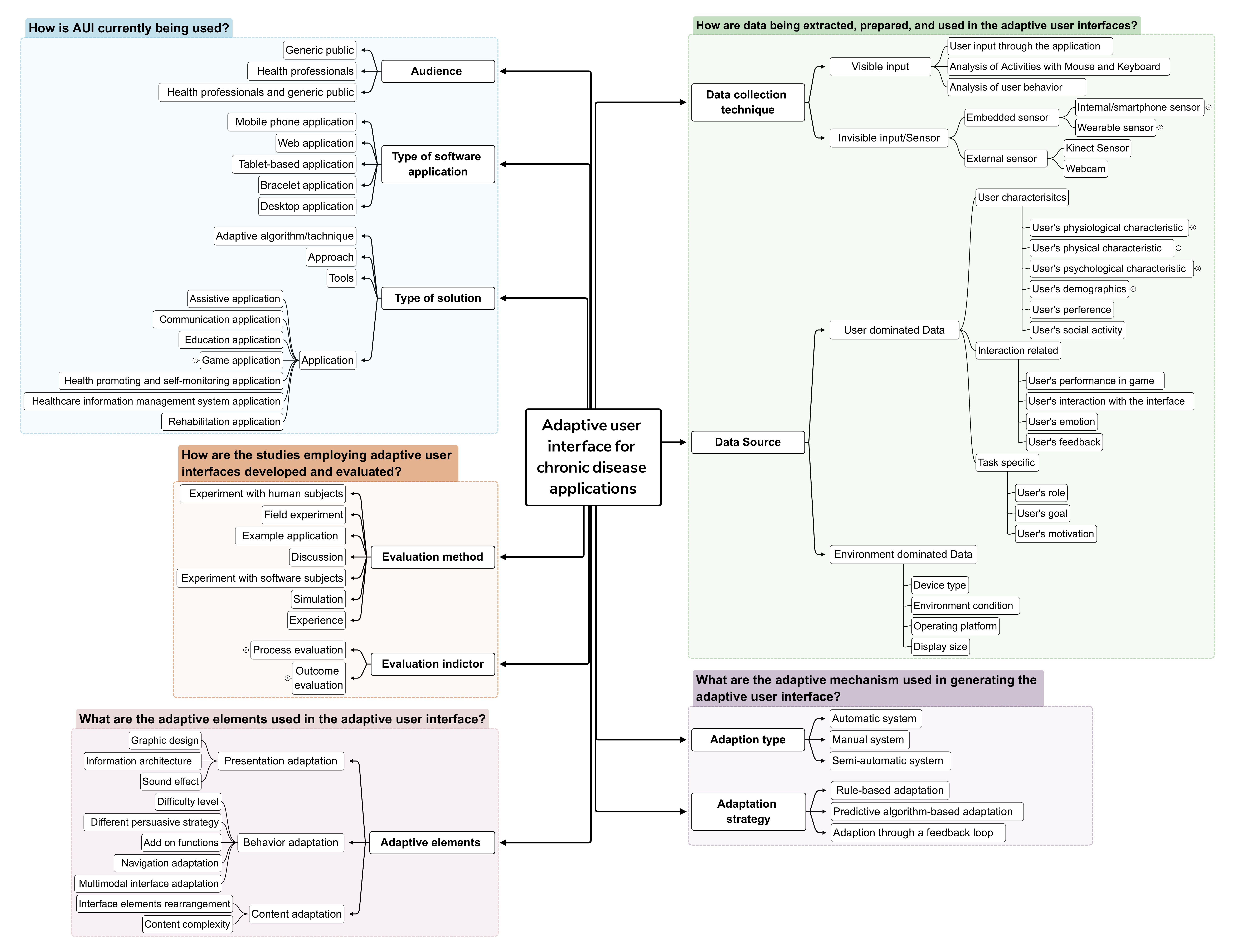}
  \caption{Mind map summary of findings}
%   \Description{Mind map summary of findings}
  \label{fig:Taxonomyall}
\end{figure}

The findings of this SLR have been presented in the previous section with respect to the RQs. We  consolidate and visualize all the findings in a mind map, as shown in Figure \ref{fig:Taxonomyall}. With these findings in mind, it is opportune to discuss the outcomes and contemplate potential avenues for future exploration and investigation.

\subsection{AUIs in various software applications} 
\begin{figure}[ht!]
  \centering
  \includegraphics[width=\linewidth]{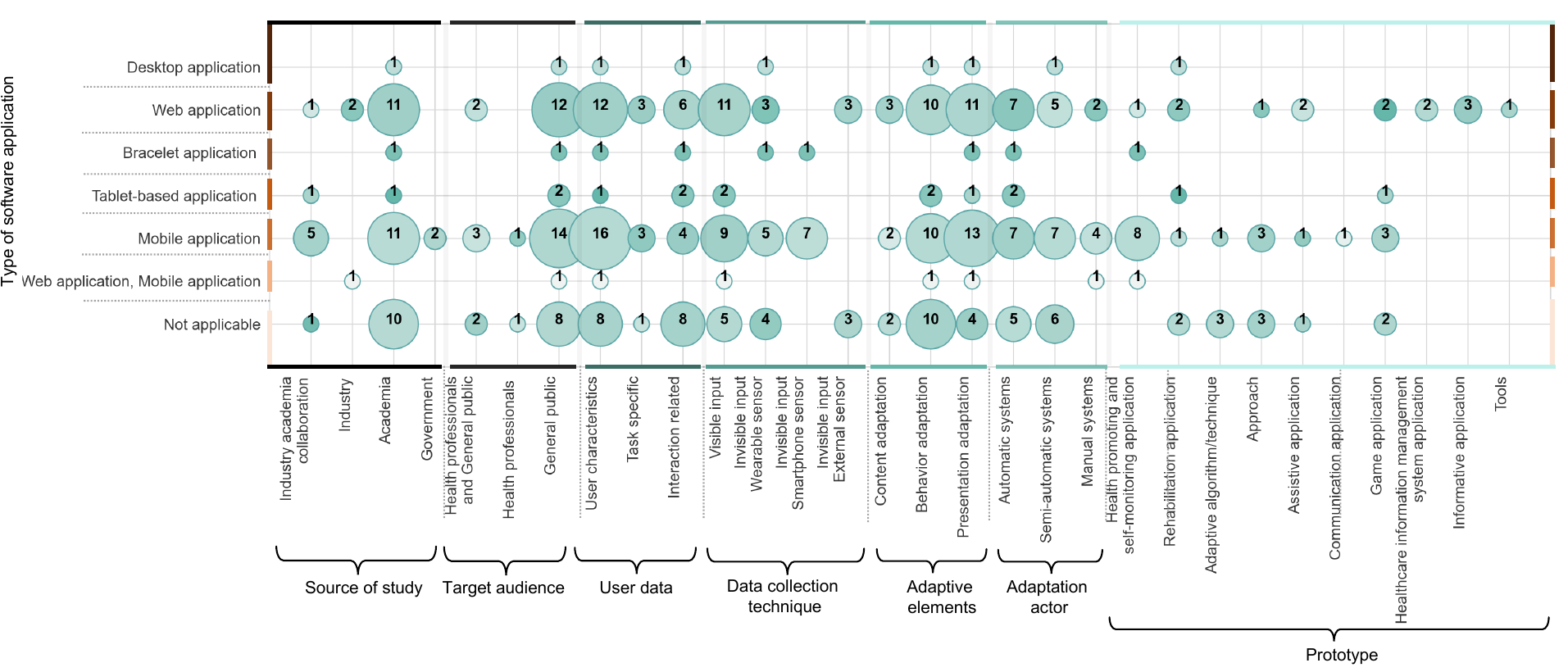}
  \caption{Bubble chart between the type of solution, source of study, target audience, user data, data collection technique, adaptive elements and adaptation actor}
  \label{fig:application}
\end{figure}

Figure \ref{fig:application} presents a mapping of the identified software types in Section \ref{sec:RQ1} into the aspects of AUI reported in Sections \ref{sec:overview}, \ref{sec:RQ2}, \ref{sec:RQ3}, \ref{sec:RQ4} and \ref{sec:RQ5}. This mapping provides a comprehensive view of the types of software associated with different aspects of AUI. We observed that \emph{web applications and mobile applications} are the primary platforms for delivering applications with AUI. Moreover, Figure \ref{fig:application} indicates that both two channels have a consistent trend for \emph{targeting users, study sources, user data, adaptive elements and adaptation actors}. 
\begin{itemize}
    \item \textbf{Targeting user groups.} \rev{We found a limited focus on adapting applications for health professionals. Even in cases where applications target both the general public and health professionals, adaptation for the latter group remains rare (e.g., S2 and S13). Typically, health professionals' involvement centres around customising training or treatment plans based on the user's health status and demographics} (e.g., S13 and S15). In future studies, there is a promising opportunity to enhance the role of AUIs in healthcare applications by addressing the specific needs of health professionals. By leveraging AUI, \emph{routine administrative and data-related tasks} could be streamlined and automated, allowing healthcare practitioners to allocate more time to direct patient care \citep{eslami2018user, Vogt2010, alnanih2012characterising}. Moreover, there is potential for AUI to play a role in supporting \emph{decision-making processes} \citep{greenwood2003agent}. \\
    \item \textbf{Varied data collection techniques.}
    Mobile applications are predominately utilised as the delivery channel for HealPM applications (See Figure \ref{fig:application}). Smartphones come equipped with a range of sensors, including acceleration sensors, vibration transducers, and angular transducers, that offer various services such as biometrics authentication, motion detection, and obstacle detection \citep{okumura2005study}. These sensors provide users with valuable information about their health, environment, activities, behaviours, and intentions \citep{mukhiya2020adaptive}. Mobile phone-based applications have proven advantageous for data capture and transfer \citep{baig2015mobile}. However, there are common technical challenges, such as battery failure and connectivity issues, that need to be addressed. For instance, the analysis and process of raw data within a mobile phone can result in notable drawbacks, primarily concerning power consumption and battery life. This scenario presents a greater challenge in comparison to the issue of low signal strength in rural areas \citep{donker2013smartphones, van2020implementation}. Participants' concerns about the relatively \emph{short battery life} of mobile phones have been highlighted in studies such as S44. Even with a battery duration of 16 hours, the requirement for constant communication with the device raises concerns among participants. Similarly, S21 reported that users have expressed concerns about increased battery consumption when utilising the multiple features of the mHealth applications for elderly monitoring, as certain features necessitate the device to remain active at all times and prevent it from entering a "deep sleep" mode. The utilisation of mobile communications, such as 4G and 5G networks, for transferring user data in smartphone-based applications represents an additional barrier. The situation is exacerbated by the \emph{high cost of mobile phone contracts and termination fees}, thereby restricting access for less well-off socio-economic groups \citep{van2020implementation, hamine2015impact, marcolino2018impact}.\\ 
    
    The \emph{sustainability of application usage} is a concern regardless of the employed data collection method \citep{baig2015mobile, van2020implementation}. It is crucial to acknowledge that the effectiveness of sophisticated data collection technology can diminish if users do not utilise the technology appropriately or inadvertently neglect to carry their phones/other wearable sensors to record the necessary information, as evidenced in studies S36 and S45. The \emph{variability of smartphone platforms} presents another obstacle in the development and implementation of smartphone-based applications, attributable to the existence of diverse operating systems. Further, the utilisation of different data collection methods raises concerns regarding compatibility between programming languages and application environments. In S21, researchers identified issues where certain devices impeded the measurement of accelerator data in stand-by or "off-screen" mode, underlining the importance of considering device compatibility. \\
    
    It is highly advisable for developers to conduct comprehensive investigations into the challenges and critical issues pertaining to the utilisation of diverse data collection techniques. By undertaking these proactive measures, developers can enhance the sustainability, reliability, and cost-effectiveness of application usage. This, in turn, will contribute to improved effectiveness and overall performance in developing AUI for various applications.
\end{itemize}

\subsection{Mapping user data}
Optimal utilisation of user data requires attaining a well-balanced alignment among multiple factors, including the specific data type, the effort required from users to input their data, and the associated costs \citep{philip2022data}. To optimise the efficacy of the AUI, it is imperative to identify \emph{a minimal set of determinants} that yield substantial impact on the desired outcome \citep{stephanidis1997decision}. This entails identifying the key factors that hold the highest value in driving significant changes in the UI. Developers can effectively determine the appropriate user data and data collection methods by following two fundamental steps. \\

\textbf{Step 1:} \emph{What is the purpose of the applications?} The specific data types relevant to the application's objectives should be identified as a priority, thereby informing the choice of an appropriate data collection method. For instance, applications that aim to provide informative content about the user's medical condition necessitate the collection of the user's disease information as well as their ability to comprehend and access various sources of information. In such cases, manual input from users emerges as the most efficient data collection approach. Conversely, health monitoring applications require the continuous collection of physiological data, which can be achieved through the utilisation of wearable or smartphone sensors \citep{shakshuki2015adaptive, Harman2014}. \\

\textbf{Step 2:} \emph{What is the primary usage context for the application?} Evaluating whether the application is predominantly utilised in a \emph{home environment, hospital setting, or daily life scenario}, enables the selection of data collection techniques that seamlessly align with users' routines. For instance, if the application is primarily used at \emph{home} (e.g., S29, S39 and S46), it may be more feasible to employ data collection techniques that leverage existing devices in the home or rely on manual input from users. This approach takes advantage of the increasing trend of smart homes, where various devices communicate with one another seamlessly. By tapping into this interconnected ecosystem, the application can gather relevant data from devices such as smart scales, fitness trackers, or home automation systems, providing a comprehensive view of the user's health and lifestyle \citep{alaa2017review}. On the other hand, if the application is intended for use in a \emph{hospital} setting (e.g., S23 and S28), more sophisticated data collection techniques and sensors may be available to monitor users' vital signs or gather specific health-related data accurately \citep{philip2022data}. In such cases, ensuring compatibility and interoperability with existing hospital systems and devices becomes crucial to avoid disruptions in the healthcare workflow and ensure the application's seamless integration into the clinical environment \citep{dinh2019wearable}. When the application is designed for use in \emph{daily life situations}, data collection techniques should be seamlessly integrated into users' routines, such as leveraging smartphone or wearable sensors, offering convenience and ease of use. By considering the context of daily life, developers can minimise the disruption to users' activities while still capturing the necessary data for the application's intended purpose. By adhering to these two steps, developers can make informed decisions regarding the appropriate user data and data collection methods.

\begin{itemize}
    \item \textbf{Type of data used.} Throughout our study, it is clear that certain types of user data and environmental data have received minimal attention in the existing literature (e.g., task-specific data). This indicates that these underrepresented data types hold great potential for future applications. In addition, we believe there is an opportunity to \emph{combine multiple data types} for a more comprehensive representation of the user. For instance, the combination of sensor data from a wearable or smartphone with contextual information like noise level, weather conditions, or social interactions, presents an opportunity to create a more comprehensive and nuanced user profile. Moreover, the combination of diverse data types can enhance the accuracy and reliability of the predictive algorithm. By incorporating multiple dimensions of information, algorithms can provide more precise and tailored insights, leading to more effective interventions and support. According to Section \ref{sec:RQ2a}, \emph{user characteristics and interaction data} are by far the area with the most effort when it comes to the data source of the adaptation and they are often used in combination with each other. For example, for user characteristics data, some studies used multiple dimensions of user characteristics to obtain a more comprehensive understanding of users' attributes, traits, and behaviours (e.g., S3, S9 and S17). There is no agreed amount of user information for the AUI, rather it depends on the purpose of the application \citep{kreuter2003tailored}. Additionally, we are able to correlate different applications with particular data types, primarily because a given data type is typically \emph{preferred} for a specific type of application \citep{KamelGhalibaf2019}. However, current research has a fundamental problem in the ability to \emph{verify and quantify} these implicit associations. Thus, there exists an open research problem in \emph{explaining which given user data was preferred to be collected as the data source of adaptation for the given application type}.\\
    
    \item \textbf{Data collection techniques.} Based on our analysis in Section \ref{sec:RQ2b}, there is a need for future studies that use adaptive technologies to make the software highly \emph{flexible} and \emph{less expensive} by properly \emph{combining software and hardware}. For instance, in the rehabilitation application, several studies commonly use Kinect sensors for human skeleton tracking (e.g., S35, S37 and S40). However, S4, one of the included studies, track virtual skeleton joints directly from the RGB data captured from a camera, which can contribute to the use of rehabilitation application more \emph{accessible} and \emph{cheaper} for a \emph{wider population}. \rev{Medical sensors, while extensively employed, can incur additional costs for individuals, potentially leading to their restricted use within hospital contexts.} In order to drive advancements in this field, future studies should prioritise the investigation and exploration of more affordable options for data collection techniques \citep{Palomares-Pecho2021,philip2022data}. Researchers can facilitate broader accessibility and affordability, thereby enabling a wider user base to benefit from the technology. The sustainability of external sensor usage constitutes another significant barrier \citep{kansal2004controlled}. In a specific study (S45), participants encountered inconveniences with the pedometers that were meant to be worn on a belt. Some participants expressed challenges in remembering to wear the device and log their activity, given the multitude of tasks they had to remember throughout the day. Consequently, the continuous sustainability of the exercise game is compromised due to these factors. Similarly, in study S36, which developed a rehabilitation game incorporating an external measurement device, users commonly encountered challenges regarding the device's usage. Forgetting to carry the device or failing to recharge its batteries are recurring issues. These oversights hindered the smooth integration of the measurement device into the rehabilitation game, thereby impacting its overall effectiveness and sustainability. To summarise, efforts should be directed towards improving the \emph{sustainability of the usage} of these data collection techniques. This entails developing solutions that address potential barriers such as user forgetfulness, user acceptability, and ease of integration into daily routines.
\end{itemize}

\subsection{Decision-making process for adaptation}
Detecting the need and making the decision to adapt the interface are pivotal stages in adaptation \citep{abrahao2021model}. According to our analysis of the results in Section \ref{sec:RQ3}, there is no \emph{one-size-fits-all} method or strategy applicable to all circumstances of adaptation decision-making.
\begin{itemize}
    \item \textbf{Adaptation strategy.} Drawing from the findings discussed in Section \ref{sec:RQ3a}, the majority of adaptive systems analysed in this SLR rely on the use of \emph{simple decision rules}. This observation is consistent with a previous review study \citep{KamelGhalibaf2019}, which highlights the common usage of rule-based approaches in health applications that involve tailored information. However, it should be noted that certain applications in our study have employed alternative algorithms, including reinforcement learning (S35), CNNs (S32), evolutionary algorithms (S14), and Monte Carlo tree search algorithm (S31), to facilitate the adaptation process.\\
    
    The selection of an adaptation strategy for mHealth applications should be based on careful consideration of several factors. \textbf{1)} It is crucial to examine the specific adaptation requirements to determine if they can be effectively captured using explicit rules. For example, in cases where there is readily available expert knowledge in a particular health domain that can be translated into rule-based conditions and actions, a rule-based adaptation approach may be suitable. Similarly, simple adaptations such as font size selection based on user age and preference, which involve straightforward adaptation logic, can also be addressed using a rule-based approach (e.g., S1, S2, and S10). However, when the adaptation requirements involve uncovering unknown patterns and analysing extensive and diverse user-generated data sets produced during application usage, an adaptation strategy based on predictive algorithms may be more appropriate \citep{langley1997machine, todi2021adapting}. \textbf{2)} Another important point to consider is the significance of transparency in the mHealth application. Transparency plays a pivotal role in fostering trust between users and mHealth applications by providing a clear understanding of how data is handled, processed, and utilised within the application \citep{wicks2015trust, sunyaev2015availability}. Rule-based adaptation offers a high level of transparency, as the adaptation process is often clear and understandable to users \citep{mukhiya2020adaptive, jokste2017rule}. This makes rule-based adaptation particularly suitable for scenarios that involve critical health-related decisions (e.g., S20), where transparency is crucial for users to have confidence in the adaptation process and the resulting outcomes. \textbf{3)} The third point to note is whether the system requires adaptation to dynamic and evolving user behaviour or environmental conditions. One downside of utilising basic if-then rules is that the developer must \emph{account for any possible variation}. The resultant adaptation may be \emph{incoherent} or \emph{sub-optimal} for the target user \citep{KamelGhalibaf2019}. In such instances, predictive algorithm-based adaptation emerges as a more suitable alternative. This approach can effectively handle the complexity and uncertainty tied to dynamic user behaviour, thereby enabling more effective adaptations.\\
    
    \item \textbf{Adaptation actor.} We have identified and discussed three different adaptation actors in Section \ref{sec:RQ3b}. \rev{The prevalence of automatic and semi-automatic systems is unsurprising, given that manual systems place the entire responsibility of adaptation on users.} However, when it comes to the manual system, there is a pressing need for a more \emph{natural adaptation} \citep{Palomares-Pecho2021}. This is crucial to overcome adoption barriers associated with the implemented AUI, including the time and effort required to learn its mechanics \citep{zhang2021designing}. For instance, S5 developed a personalisation rule editor that enables users to specify their own adaptation rules based on contextual elements, triggers, and actions. However, this approach may pose challenges for users with limited computer literacy, particularly in grasping the complexities of triggers and actions, which often involve elements akin to programming language. \\
    
    Our review has also revealed that a simple rule-based adaptation can \emph{distribute the initiative} (what initiates the adaptation process) between the user and the system \citep{abrahao2021model}. \rev{This implies that either the user can manually adjust variables to different settings or the system can autonomously carry out the adaptation process based on predefined rules.} It highlights the existence of varying levels of \emph{user involvement} in the adaptation process, which can be influenced by factors such as the \emph{user's willingness} to drive the adaptation process and their level of \emph{knowledge} required for making adaptations \citep{eiband2021support}. Consequently, there is an important need for research to \emph{gain a deep understanding of distributing the initiative between the end-user and the system}. Understanding these dynamics is crucial for designing AUI that cater to users' preferences and capabilities, while also considering their comfort level in taking charge of the adaptation process. By recognising and accommodating different levels of user involvement, developers can create AUI that strikes a balance between user empowerment and system support, resulting in a more personalised and satisfactory user experience \citep{eiband2021support}. This motivates the following questions: \emph{What level of automation do users prefer for different types of applications or adaptive elements? How can we control the user's involvement in the whole adaptation process (e.g., who triggers the adaptation or evaluates the adaptation results)?} 
\end{itemize}

\subsection{Adaptive elements}
Through our analysis conducted in Section \ref{sec:RQ4}, \rev{it is clear that certain types of adaptive elements receive minimal attention in the literature (e.g. content complexity, interface element arrangement), while graphic design and difficulty level stand out as the most frequently utilised among the included studies. Moreover, it is notable that the adaptation process usually involves the modification of multiple elements within} the software. This characteristic stands out and presents a crucial avenue for future research: \emph{what is the combination of adaptation elements that users are satisfied with?}. To address this, it is crucial to place the user at the core of the adaptation process. By conducting comprehensive user research, gathering feedback, and actively involving users in the design and evaluation phases, researchers can gain deep insights into users' specific needs and preferences regarding how they desire different adaptive elements to be employed in conjunction with one another. However, it is worth noting that only a limited number of included studies have investigated user needs during the initial phases of AUI development (as discussed in Section \ref{sec:dissdesign}). This underlines the necessity for a more comprehensive user-centred approach that accounts for user preferences across different adaptive elements.

Another important issue is that applications designed to promote physical activity and rehabilitation exercises always seem to provide adaptations to behavioural levels and will also incorporate adaptations using graphic design at the same time. Researchers can further analyse \emph{which types of adaptations users prefer}, particularly in relation to the \emph{purpose of these applications}. The response to this question may vary based on the user. Thus, \emph{examining user types and motives} could constitute a compelling area for future research within this domain. For example, in rehabilitation exercises, it could be valuable to investigate whether users are more motivated by adjustments to the speed of an animated trainer and other factors tied to the difficulty level, as opposed to being incentivised by rewards for their accomplishments. Examining how different types of users respond to various adaptation strategies and identifying the most effective approaches for different user groups and their objectives would contribute to the refinement and optimisation of AUI design in these applications.

\subsection{System design for the AUI}\label{sec:dissdesign}
Based on the results from Section \ref{sec:RQ5}, the general \emph{lack of information} on system design (69\% missing) suggests serious deficiencies in the reporting of design methodology in the literature on AUI related to chronic disease management software. As reported in Section \ref{sec:RQ5}, descriptions of eHealth interventions are often restricted to \emph{focusing on enhanced user experiences} and their \emph{efficiency in improving health outcomes and behaviours} (evaluation results). Few of our selected studies explain how the proposed solutions were developed, especially in the early stages of design stages when end users were involved (e.g., S5 and S6). Even with studies that mentioned how they develop the system, very little information was provided without any further description (e.g., S12, S44 and S46). Therefore, we have to abandon them for interpretation and exclude the design approach in the taxonomy (Figure \ref{fig:Taxonomyall}). Two previous studies have criticised the lack of information about the design process \citep{KamelGhalibaf2019,dabbs2009user}. Therefore, we strongly recommend more \emph{coverage of the system design process} for reported studies in future research. Reporting of the design process not only improves the \emph{understandability} of the proposed solution but also allows readers to better assess the \emph{validity and reliability} of the reported approaches, tools and applications with AUI. 

\subsection{System evaluation for AUI}
The SE community has long emphasised the use of \emph{scientific and rigorous} evaluation methods to assess the software \citep{Hachey2012,Zannier2006}. According to the results reported in Section \ref{sec:RQ5}, the findings and opportunities for future research are as follows:
\begin{itemize}
    \item \textbf{Length of evaluation time and participants number.} A key finding from Section \ref{sec:RQ5} is that a considerable number of the included studies were evaluated with a \emph{small sample} and \emph{short evaluation period}. This concern is of higher relevance in the context of chronic disease \citep{abowd1999towards}. A major reason for this higher relevance is that the effectiveness of chronic disease prevention and rehabilitation applications will require \emph{longitudinal} studies \citep{Palomares-Pecho2021}. In addition, there is also a greater need for a large number of participants due to the large \emph{variations} between individuals, which is the basis for adaptive technology \citep{Norcio1989, Akiki2014}. \rev{Therefore, we highly recommend that future investigations into AUI development for chronic disease applications prioritise the inclusion of \emph{larger participant samples}  and have \emph{longer evaluation duration}}.\\
    \item \textbf{Adaptation evaluation.} Based on evaluation analysis in Section \ref{sec:RQ5}, the most common method of evaluation is to simply test the effectiveness (e.g., user experience, behaviour outcome of the users) of the application as a whole, making it difficult to draw conclusions regarding the AUI implemented. Our analysis of the evaluation results revealed a notable \emph{inconsistency} in assessing the quality of the adaptation. There are few or no explicit recommendations on how to evaluate the adaptation in order to support the early phase of designing AUI \citep{abrahao2021model}. There are some existing criteria to evaluate the adaptation, such as \textbf{\emph{specification of adaptation}} (\textit{"refers to the user's ability to specify the actions required to make this adaptation"}) \citep{lopez2007towards} and \textbf{\emph{user feedback of adaptation}} (\textit{"refers to the ability of the approach to provide feedback about the quality of the adaptation."}) \citep{abrahao2021model}, but their indicator and usage are not made sufficiently explicit to be adopted by developers for evaluating the adaptation. Thus, there is an opportunity for researchers to develop guidelines and establish \emph{adaptation evaluation infrastructure},  encompassing metric development and facilitating comparative analysis of adaptations. Such work would allow for clearer and more concise evaluations of adaptation, solidifying claims made from the results of a given evaluation. 
\end{itemize}
\section{Threats to validity}\label{sec:threat}
Even though this SLR was carried out following a well-established approach \citep{Kitchenham2007,kitchenham2022segress}, our review still has some limitations, most of which are related to our search methodology and the data extraction process.

\textbf{Construct validity.}
The extent to which the research reflects the researcher's intention and what is investigated by the RQs \citep{Wohlin2012}. 
The inadequacy of the search and study selection process is the most evident bias that might compromise the construct validity of this study \citep{Shahin2014}. For example, although the concept of the AUI is well known, systems that employ it may also be referred to by terms like "smart user interface" or "intelligent user interface". To mitigate any possible threats in the search strategy, we employed two strategies: \textbf{1)} We concluded that missing such terms posed a negligible risk in this SLR after doing several rounds of trial searches in six well-known digital libraries. \textbf{2)} We consulted the search strings utilised in previous SLRs \citep{Goncalves2019, Aranha2021} and used the PICOC criteria to ensure that the search is thorough and comprehensive. As a result, our search approach works well enough to identify significant papers in the related field. Another threat to construct validity lies in the search of paper metadata (i.e., titles, abstracts, and keywords) in most databases, \rev{potentially leading to the systematic rejection of publications lacking validated references in their metadata.}

\textbf{Internal validity.}
The extent to which the design and execution of our SLR study are likely to avoid systemic errors \citep{Wohlin2012}. The derived taxonomy characterises the field of AUI designed for chronic disease-related applications, as the key contribution of this study. To mitigate any errors in the proposed taxonomy, we took three strategies: \textbf{1)} We developed a data extraction form to gather and analyse data consistently in order to answer the SLR's RQs. \textbf{2)} Four authors extract the same number of the included papers independently. \rev{A comparison of the extracted data was undertaken to identify and rectify any contradictory information.} \textbf{3)} We followed the open coding process of constructivist grounded theory \citep{charmaz2014constructing}. In our SLR, each attribute classification was reviewed and refined by at least three authors until all three agreed. To boost the integrity of our taxonomy and increase the transparency of the data extraction process, all data extraction details are included in the supplementary material so that the reader can confirm the reliability of the information extracted. 

\textbf{External validity.}
The extent to which the original studies are representative of the reviewed field may result from selection bias \citep{Wohlin2012}. To mitigate this bias, we implemented the following three steps: \textbf{1)} We strictly executed this SLR according to the SLR protocol. The SLR protocol was arranged by one author and subsequently refined by other authors. The initial study set was refined using inclusion and exclusion criteria such that only studies that meet the scope were included. \textbf{2)} Any disagreements that surfaced during the study selection process were resolved during the internal discussion. After the conversation, we also documented the reasons for inclusion or exclusion. \textbf{3)} We also used snowballing techniques to find as many relevant papers as possible to reduce selection bias.

\textbf{Conclusion Validity.}
Potential biases regarding the existence or absence of relationships may lead to incorrect conclusions \citep{Wohlin2012}. To mitigate this threat, the data extraction results were plotted and correlated using various graphs in addition to textual descriptions. This helps to enhance the traceability between the extracted data and the conclusions.

\section{Conclusion}\label{sec:conclusion}
We presented a Systematic Literature Review of primary studies related to AUIs as used for chronic disease-related software applications. Our work used the guidelines laid out by \citep{Kitchenham2007,kitchenham2022segress} for performing SLRs in SE. We generated a taxonomy that pertains to different aspects of applying AUIs to chronic disease-related applications. Our analysis of 48 included primary studies allows concluding that the most used source of data for the adaptation, data collection techniques, the decision-making process of the adaptation and the final action made to the interface. The concepts described in this review should aid researchers and developers in understanding where AUI can be applied and the necessary considerations for employing AUIs in different chronic disease-related applications. Key future research directions include: \textbf{(1)} \rev{Investigating the involvement of health professionals in the adaptation process and how to benefit health professionals by boosting their efficiency within their daily responsibilities. \textbf{(2)} Exploring users' preferences concerning the collection of data as a foundation for the adaptation process. \textbf{(3)} Exploring innovative, lower-cost and effective options of data collection techniques. \textbf{(4)} Exploring combinations of different adaptive strategies and levels of automation users prefer for given types of applications or adaptive elements. \textbf{(5)} Examining users' preferred adaptive elements, while considering the unique purpose of different applications. \textbf{(6)} Adhering to established standard reporting guidelines for original research development and formulating effective guidelines for evaluating adaptation approaches within the realm of chronic disease applications.}

\bmhead{Supplementary information}

We provide additional information that supports our findings and SLR process, including the SLR planning process, SLR selection process, SLR data extraction process and Data analysis.

\bmhead{Acknowledgments}

Wang, Madugala and Grundy are supported by ARC Laureate Fellowship FL190100035

\section*{Declarations}
Not applicable

\bigskip




\begin{appendices}

\section{Research Methods}
\subsection{Data extraction form}\label{app:dataeextraction}
{\tiny
    \begin{longtable}{p{1mm}p{45mm}p{50mm}p{6mm}}
    \caption{Data extraction form}\\
      \hline
        \textbf{\#}&\textbf{Questions}&\textbf{Description/examples}&\textbf{RQs}\\
      \hline
      \endfirsthead
      
      \multicolumn{4}{l}%
        {{\bfseries \tablename\ \thetable{} -- continued from previous page}} \\\hline
        \textbf{\#}&\textbf{Questions}&\textbf{Description/examples}&\textbf{RQs}\\
     \hline
    \endhead
    
    \hline
    \multicolumn{4}{r}{{Continued on next page}} \\ \hline
    \endfoot
    
    \hline
    \endlastfoot
    
    1&Paper ID&\multicolumn{2}{l}{\multirow{13}{*}{Demographic data}} \\
    2&Paper Title&\multicolumn{2}{l}{}\\
    3&Authors of the Paper&\multicolumn{2}{l}{}\\
    4&Venue &\multicolumn{2}{l}{}\\
    5&First Authors affiliated university/institution  &\multicolumn{2}{l}{}\\
    6&First Author affiliated country&\multicolumn{2}{l}{}\\
    7&The study population affiliated country in the paper&\multicolumn{2}{l}{}\\
    8&Type of publication&\multicolumn{2}{l}{}\\
    9&Publication Year &\multicolumn{2}{l}{}\\
    10&Source Type&\multicolumn{2}{l}{}\\
    11&Citations number&\multicolumn{2}{l}{}\\
    12&Page number&\multicolumn{2}{l}{}\\
    13&Study context&\multicolumn{2}{l}{}\\
    \hline
    14& What are the major keywords of the study? &\multicolumn{2}{l}{\multirow{2}{*}{Information for paper}} \\
    15& What is the aim/ motivation/ goal of the study?&\multicolumn{2}{l}{}\\
    \hline
    16& What type of software application has been presented in the studies?& e.g., web applications, mobile application, desktop applications or tablet applications.& RQ1\\
    17&What type of solution is offered in the studies?&e.g., architecture, framework, platform, algorithms and application prototypes.&RQ1\\
    18& What health conditions have been targeted in the studies? &e.g., heart disease, cardiovascular disease, diabetes, hypertension, alzheimer and pulmonary disease. & RQ1\\
    19& What is the target audience of the paper?& e.g., patients and health professionals.&RQ1\\
    20&Is there a specific aged group if the paper is targeted at end-users?&specify the age group.&RQ1\\
    21&What is the solution adapt to?&specify the user group the solution can adapt to.&RQ1\\
    \hline
    22&What environment data is collected to generate AUI for the application?&e.g., different mobile devices, change of the light, change of location and different operating platform.&RQ\textsubscript{2a}\\
    23&What user-controlled input data is collected to generate AUI for the application?& e.g., user's performance when using the app, user's emotions are address, user’s personality, user’s disease, user’s disabilities, user’s age and user’s heart rate.&RQ\textsubscript{2a}\\
    24&What techniques are being used to collect the data?& e.g., Electrocardiogram (ECG), Electrodermal activity (EDA), Ectroencephalogram (EEG), Respiration Analysis and Electrooculography (EOG) measuring eye movements. &RQ\textsubscript{2b}\\
    \hline
    25&What are the adaptive strategies used in generating the AUI?& technique used to tailored the user interface, e.g. rule-based adaption, predictive algorithm-based adaption and goal-driven adaption (Try to put as much details as you can, if you are not sure which adaption strategies the study is adopted).&RQ\textsubscript{3a}\\
    26& Who is adapting? & e.g., automatic systems( self-adaptation).semi-automatic systems (by algorithms and user). manual system (user manually adapts).&RQ\textsubscript{3b}\\
    \hline
    27&What are the adaptive elements used in the AUI for applications?&what component is changed on the interface (Try to put as much details as you can, if you are not sure of the elements).&RQ\textsubscript{4}\\
    \hline
    28&What are the approaches (e.g., frameworks, environments, design methodology etc.) to develop applications with AUI?&e.g., model-based approach,  model-based user interface development approach and multi-agent architecture (Try to put as much details as you can, if you are not sure the approach).&RQ\textsubscript{5}\\
   29&What are the types of models used in applications with the AUI?&e.g., context model, dialogue model, task model, domain model, user model, etc.&RQ\textsubscript{5}\\
    30&Main outcome/ Results of the study?&what did this study propose? Does the proposed solution usable, effective?&RQ\textsubscript{5}\\
    31&How do they evaluate their results?&&RQ\textsubscript{5}\\
    32&What approaches are used to evaluate the solutions?
    &we adopted the \emph{classification scheme} for evaluation approaches from \cite{Chen2011}. e.g., case study, discussion, example application, experience, and experiment with human subject.&RQ\textsubscript{5}\\
    33&How the evaluation results being measured?&e.g., behavioural outcome, health/medical status, user experience, psycho-behavioural determinant, knowledge level, and usability metric. &RQ\textsubscript{5}\\
    34&What is the length of time for the evaluation?
     & e.g., how many session they have for the evaluation, how long for each session, will they follow up?&RQ\textsubscript{5}\\
    35&What are the major recommendations of the study?&\multirow{3}{*}{key research gaps/future work/limitations}&RQ\textsubscript{5}\\
    36&What are the main limitations and main strengths of the study?&&RQ\textsubscript{5}\\
    37&What are the key research gaps/ future work identified by each study?&&RQ\textsubscript{5}\\
    \end{longtable}
}
\clearpage
\subsection{Quality assessment}\label{app:qualityass}
{\footnotesize
    \begin{longtable}{p{5mm}p{105mm}}
    \caption{Study quality assessment questions}\\
    \hline
    \textbf{\#}&\textbf{Quality assessment questions}\\
     \hline
     \endfirsthead

    \multicolumn{2}{l}%
    {{\bfseries \tablename\ \thetable{} -- continued from previous page}} \\
    \hline
    \textbf{\#}&\textbf{Quality assessment questions}\\
    \hline
    \endhead
    
    \hline \multicolumn{2}{r}{{Continued on next page}} \\ \hline
    \endfoot
    \hline
    \endlastfoot
    
    \textbf{QA1}& Are the aims and objectives of the study clearly specified?\\ 
    \textbf{QA2}& Is there an adequate description of the context in which the research was carried out?\\
    \textbf{QA3}& Does the research design match the aims claimed for the research?\\
    \textbf{QA4}& Does the paper present a detailed description of the AUI employed?\\
    \textbf{QA5}& Is there a clear outcome and results analysis reported?\\
    \textbf{QA6}& Does the paper provide limitations, summary and future work of the research?\\
    \textbf{QA7\textsubscript{a}}& What is the quality of the venue where the study was published? Rated by considering the CORE 2021 \citep{ConferencePortal2021} for the conference papers and ranking of conference and Journal Citation Reports 2021 \citep{EugeneG2022} for the rankings of journal papers.\\
    \textbf{QA7\textsubscript{b}}& What is the total number of citations for the paper?
    \label{tab:qualityassessment}
    \end{longtable}
    }

\subsection{Data Synthesis, and Taxonomy Derivation}\label{app:datasynthesis}
We downloaded all the essential information in a data extraction sheet, including \textbf{a)} demographic data (e.g., title, authors, venue, affiliation and publication type), \textbf{b)} the answers for each RQ, \textbf{c)} the QA scores for all QA questions. Both qualitative and quantitative methods were used to synthesise the extracted data. Additional data for the data analysis are given in supplementary material.

\textbf{Quantitative analysis}: We performed both \emph{univariate and multivariate frequency distribution analysis}. The \emph{univariate} frequency distribution offered a summary count of the occurrences within a particular variable. \emph{Multivariate} frequency analysis was used to aggregate the distribution of two or more variables to determine their interrelationships. We used Python and Microsoft Excel pivot table tools to build and visualise the cross-tabulation and other diagrams.
\begin{figure}[hb]
%   \centering
  \includegraphics[width=\linewidth]{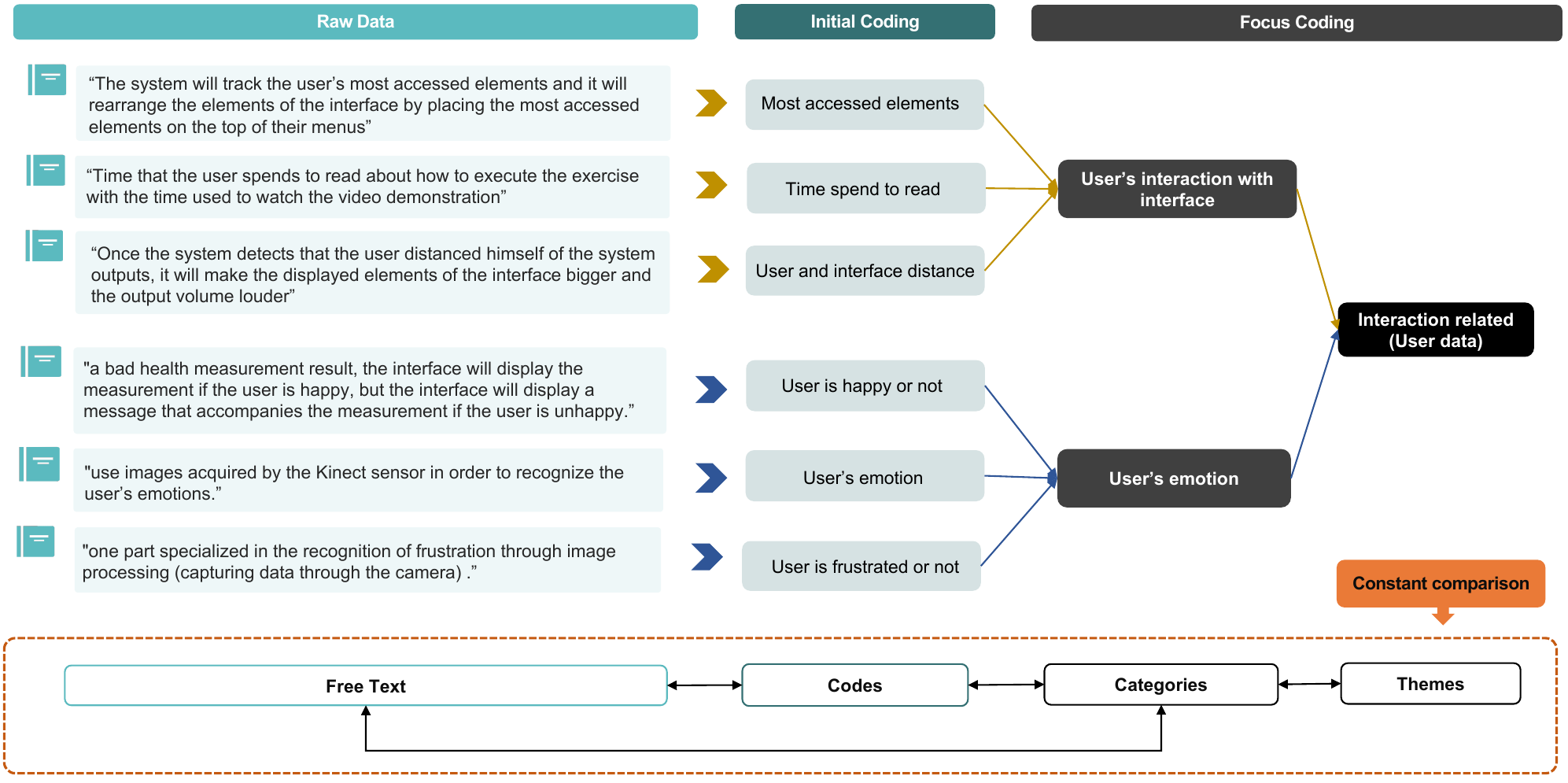}
  \caption{Example of qualitative data analysis by using constructivist grounded theory}
%   \Description{Example of qualitative data analysis by using constructivist grounded theory}
  \label{fig:groundedthe}
\end{figure}
\textbf{Qualitative analysis}: To build our taxonomy on the use of AUI, we followed an \emph{open coding methodology} consistent with \emph{constructivist grounded theory} \citep{charmaz2014constructing}. According to the advice of recent work within the SE community \citep{stol2016grounded}, we followed the two steps as detailed below: \textbf{1)} we created the initial coding of data type, data collection technique and adaptive elements and \textbf{2)} we selected categories from the most frequent or important codes we created and used them to categorise the data (focused coding). The initial coding was performed by one author and then refined during the focused coding process by two others until an agreement was made among all three. The results of these coding steps formed our taxonomy. This coding process is visualised in Figure \ref{fig:groundedthe}. Additional data for the taxonomy are given in supplementary material.

\section{Included studies}
\subsection{List of Primary Studies}\label{appe:includedstu}

%\begin{table}[!h]
%\afterpage{
{ \tiny 
    \begin{longtable}{p{2mm} p{35mm}p{30mm}p{25mm}p{5mm}}
  \caption{List of review articles}\\
  \hline
     \textbf{ID}&\textbf{Title}&\textbf{Author(s)}&\textbf{Venue}&\textbf{Year}\\
 \hline
\endfirsthead

\multicolumn{5}{l}%
{{\bfseries \tablename\ \thetable{} -- continued from previous page}} \\\hline
 \textbf{ID}&\textbf{Title}&\textbf{Author(s)}&\textbf{Venue}&\textbf{Year}\\
     \hline
    \endhead
    
    \hline
    \multicolumn{5}{r}{{Continued on next page}} \\ \hline
    \endfoot
    
    \hline
    \endlastfoot

S1&Tools for adaptation of a mobile application to the needs of users with cognitive impairments&Dmytro Fedasyuk, Illia Lutsyk&16th International Conference on Computer Sciences and Information Technologies&2021\\
S2&Pervasive multimedia for autism intervention&S. Venkatesh, S. Greenhill, D. Phunga, B. Adamsb, T. Duonga&Pervasive and Mobile Computing&2012\\
S3&A Modular Mobile Exergaming System With an Adaptive Behavior&Ali Karime, Basim Hafidh, Wail Gueaieb, Abdulmotaleb El Saddik&IEEE Instrumentation and Measurement Society&2015\\
S4&Optimising engagement for stroke rehabilitation using serious games&J. W. Burke, M. D. J. McNeill, D. K. Charles, P. J. Morrow, J. H. Crosbie, and S. M. McDonough& Vis. Comput.&2009\\
S5&Enabling Personalisation of Remote Elderly Assistant Applications&Chesta, Cristina
Corcella, Luca, Kroll, Stefan, Manca, Marco, Nuss, Julia, Paternò, Fabio, Santoro, Carmen&ACM International Conference Proceeding Series&2017\\
S6&A Personalized Physical Activity Coaching App for Breast Cancer Survivors: Design Process and Early Prototype Testing&Monteiro-Guerra, Francisco, Signorelli, Gabriel Ruiz, Tadas, Shreya, Zubiete, Enrique Dorronzoro, Romero, Octavio Rivera, Fernandez-Luque, Luis
Caulfield, Brian&JMIR mHealth and uHealth&2020\\
S7&Experience of Designing and Deploying a Tablet Game for People with Dementia&Westphal, Bree J., Lee, Hyowon, Cheung, Ngai Man, Teo, Chor Guan, Leong, Wei Kiat&OZCHI '17: Proceedings of the 29th Australian Conference on Computer-Human Interaction&2017\\
S8&Move\&Learn: an Adaptive Exergame to Support Visual-Motor Skills of Children with Neurodevelopmental Disorders&Raygoza-Romero, Joan, Gonzalez-Hernandez, Ariana, Bermudez, Karina, Martinez-Garcia, Ana I., Caro, Karina&GoodIT '21: Proceedings of the Conference on Information Technology for Social Good&2021\\
S9&BeWell+: Multi-dimensional Wellbeing Monitoring with Community-guided User Feedback and Energy Optimization & Lin, Mu, Lane, Nicholas D., Mohammod, Mashfiqui, Yang, Xiaochao, Lu, Hong, Cardone, Giuseppe, Ali, Shahid, Doryab, Afsaneh, Berke, Ethan, Campbell, Andrew T., Choudhury, Tanzeem&WH '12: Proceedings of the conference on Wireless Health&2012\\
S10&Improving the Efficacy of Games for Change Using Personalization Models&Orji, Rita
Mandryk, Regan L., Vassileva, Julita&ACM Transactions on Computer-Human Interaction&2017\\
S11&A Mobile Rehabilitation Application for the Remote Monitoring of Cardiac Patients after a Heart Attack or a Coronary Bypass Surgery&Gay, Valérie, Leijdekkers, Peter, Barin, Edward&PETRA '09: Proceedings of the 2nd International Conference on PErvasive Technologies Related to Assistive Environments&2009\\
S12&Designing Context Aware User Interfaces for Online Exercise Training Supervision&Klompmaker, Florian, Nebe, Karsten, Busch, Clemens, Willemsen, Detlev&2nd Conference on Human System Interactions&2009\\
S13&Mobile@Old – An Assistive Platform for Maintaining a Healthy Lifestyle for Elderly People&Awada, Imad Alex, Mocanu, Irina, Jecan, Sergiu, Rusu, Lucia, Florea, Adina Magda, Cramariuc, Oana&The 6th IEEE International Conference on E-Health and Bioengineering&2017\\
S14&Dynamic Difficulty Adjustment with Evolutionary Algorithm in Games for Rehabilitation Robotics&Andrade, Kleber De O., Pasqual, Thales B., Caurin, Glauco A.P., Crocomo, Marcio K.& 2016 IEEE International Conference on Serious Games and Applications for Health (SeGAH)&2016\\
S15&Personal health monitoring with Android based mobile devices&Kozlovszky, M.
Bartalis, L., Jokai, B., Ferenczi, J., Bogdanov, P., Meixner, Zs, Nemeth, L., Karoczkai, K.&2013 36th International Convention on Information and Communication Technology, Electronics and Microelectronics (MIPRO)&2013\\
S16&A customizable mobile tool for supporting health behavior interventions&Koskinen, Esa, Salminen, Jukka&2018 IEEE 5th International Congress on Information Science and Technology (CiSt)z&2007\\
S17&A tailored, interactive health communication application for patients with type 2 diabetes: study protocol of a randomised controlled trial&Weymann, Nina, Härter, Martin, Dirmaier, Jörg&JOURNAL OF MEDICAL INTERNET RESEARCH&2015\\
S18&MediNet: Personalizing the Self-Care Process for Patients withDiabetes and Cardiovascular Disease Using Mobile Telephony&Permanand Mohan, Dylan Marin, Salys Sultan, Ahad Deen&2008 30th Annual International Conference of the IEEE Engineering in Medicine and Biology Society&2008\\
S19&PEGASO: A Personalised and Motivational ICT System to Empower Adolescents Towards Healthy Lifestyles&Carrino, Stefano, Caon, Maurizio, Angelini, Leonardo, Mugellini, Elena, Abou Khaled, Omar, Orte, Silvia, Vargiu, Eloisa, Coulson, Neil, Serrano, José C.E., Tabozzi, Sarah, Lafortuna, Claudio, Rizzo, Giovanna&Innovation in Medicine and Healthcare &2014\\
S20&Intelligent interaction interface for medical emergencies:Application to mobile hypoglycemia management&Pagiatakis, Catherine, Rivest-Hénault, David, Roy, David, Thibault, Francis, Jiang, Di&Smart Health&2020\\
S21&Protege: A mobile health application for the elder-caregiver monitoring paradigm&Ferreira, Fábio, Dias, Flávio, Braz, João, Santos, Ricardo, Nascimento, Roberto, Ferreira, Carlos, Martinho, Ricardo&Procedia Technology&2013\\
S22&User-tuned Content Customization for Children with Autism Spectrum Disorders&Da Silva, Margarida Lucas, Gonçalves, Daniel, Silva, Hugo&Procedia Computer Science&2014\\
S23&Improving Mobile Device Interaction for Parkinson’s Disease Patients via PD-Helper&Jabeen, Farzana, Tao, Linmi, Guo, Yirou, Zhang, Shiyu, Mei, Shanshan&Computers in Biology and Medicine&2019\\
S24&generating context-awareness interface for medical applications&Kallel, Fahmi, Ellouze, Afef Samet, Bouaziz, Rafik&DESE '11: Proceedings of the 2011 Developments in E-systems Engineering&2011\\
S25&Adaptive User Interface for Healthcare Application for People with Dementia&Awada, Imad Alex, Mocanu, Irina, Nastac, Dumitru Iulian, Benta, Dan, Radu, Serban&2018 17th RoEduNet Conference: Networking in Education and Research (RoEduNet)&2018\\
S26&Architecture of a System for Stimulating Intellectual Activity with Adaptive Environment SMILE&Gusev, Marjan, Tasic, Jurij, Patel, Shushma&2017 25th Telecommunication Forum (TELFOR)&2017\\
S27&Therapeutic Games’ Difficulty Adaptation: An Approach Based on Player’s Ability and Motivation&Hocine, Nadia, Gouaich, Abdelkader&16th International Conference on Computer Games&2011\\
S28&Empirical Evaluation of Intelligent Mobile User Interfaces in Healthcare&Alnanih, Reem, Ormandjieva, Olga, Radhakrishnan, Thiruvengadam&Canadian Conference on Artificial Intelligence&2014\\
S29&Computerized decision support for beneficial home-based exercise rehabilitation in patients with cardiovascular disease&Triantafyllidis, Andreas, Filos, Dimitris, Buys, Roselien, Claes, Jomme, Cornelissen, Véronique, Kouidi, Evangelia, Chatzitofis, Anargyros, Zarpalas, Dimitris, Daras, Petros, Walsh, Deirdre, Woods, Catherine, Moran, Kieran, Maglaveras, Nicos
Chouvarda, Ioanna&Computer Methods and Programs in Biomedicine&2018\\
S30&Reconfiguration of Graphical User Interface&Periyasamy, Kasi
Perkinian, Vinoth&Journal of Digital Information Management&2011\\
S31&Adaptation in serious games for upper-limb rehabilitation: an approach to improve training outcomes&Hocine, Nadia, Gouaïch, Abdelkader, Cerri, Stefano A., Mottet, Denis, Froger, Jérome, Laffont, Isabelle&User Modeling and User-Adapted Interaction&2015\\
S32&EMERALD—ExerciseMonitoring Emotional Assistant&Rincon, Jaime A., Costa, Angelo, Carrascosa, Carlos, Novais, Paulo, Julian, Vicente&Sensors (Switzerland)&2019\\
S33&Nourish Your Tree! Developing a Persuasive Exergame for Promoting Physical Activity Among Adults&Oyebode, Oladapo, Maurya, Devansh, Orji, Rita&2020 IEEE 8th International Conference on Serious Games and Applications for Health, SeGAH 2020&2020\\
S34&Adaptive Strategy for Multi-User Robotic Rehabilitation Games&Caurin, Glauco A.P., Siqueira, Adriano A.G., Andrade, Kleber O., Joaquim, Ricardo C., Krebs, Hermano I.&2011 Annual International Conference of the IEEE Engineering in Medicine and Biology Society&2011\\
S35& MPRL: Multiple-Periodic Reinforcement Learning for Difficulty Adjustment in Rehabilitation Games&Yoones A. Sekhavat&2017 IEEE 5th International Conference on Serious Games and Applications for Health (SeGAH)&2017\\
S36&Move2Play: An Innovative Approach to Encouraging People to Be More Physically Active&Bielik, Pavol, Tomlein, Michal, Krátky, Peter, Mitrík, Štefan, Barla, Michal, Bieliková, Mária&IHI '12: Proceedings of the 2nd ACM SIGHIT International Health Informatics Symposium&2012\\
S37&Enhancing the Physical Activity of Older Adults Based on User Profiles&Awada, Imad Alex, Mocanu, Irina, Florea, Adina Magda, Rusu, Lucia, Arba, Raluca, Cramariuc, Bogdan&017 16th RoEduNet Conference: Networking in Education and Research (RoEduNet)&2017\\
S38& Digital-Pheromone Based Difficulty Adaptation in Post- Stroke Therapeutic Games&Gouaïch, Abdelkader, Hocine, Nadia, Van Dokkum, Liesjet, Mottet, Denis&IHI '12: Proceedings of the 2nd ACM SIGHIT International Health Informatics Symposium&2012\\
S39&Self-Adaptive Games for Rehabilitation at Home&Pirovano, Michele, Mainetti, Renato, Baud-Bovy, Gabriel, Lanzi, Pier Luca, Borghese, Nunzio Alberto&2012 IEEE Conference on Computational Intelligence and Games, &2012\\
S40&Adaptive Gameplay and Difficulty Adjustment in a Gamified Upper-Limb Rehabilitation&Pinto, Joana F., Carvalho, Henrique R., Chambel, Gonçalo R.R., Ramiro, João
Gonçalves, Afonso&2018 IEEE 6th International Conference on Serious Games and Applications for Health (SeGAH)&2018\\
S41&Usability of an Adaptive Computer Assistant that Improves Self-care and Health Literacy of Older Adults&Blanson Henkemans, O. A., Rogers, W. A., Fisk, A. D., Neerincx, M. A., Lindenberg, J., Van Der Mast, C. A.P.G.&Methods of Information in Medicine&2008\\
S42&Adaptation of Graphics and Gameplay in Fitness Games by Exploiting Motion and Physiological Sensors&Buttussi, Fabio, Chittaro, Luca, Ranon, Roberto, Verona, Alessandro&International Symposium on Smart Graphics&2007\\
S43&Framework for personalized and adaptive game-based training programs in health sport&Hardy, Sandro, Dutz, Tim, Wiemeyer, Josef, Göbel, Stefan, Steinmetz, Ralf&Multimedia Tools and Applications&2014\\
S44&Flowers or a robot army? (encouraging awareness and activity with personal, mobile displays)&Legrand, Louis, Consolvo, Sunny, Klasnja, Predrag, Mcdonald, David W, Avrahami, Daniel, Froehlich, Jon, Libby, Ryan, Mosher, Keith, Landay, James A&UbiComp '08: Proceedings of the 10th international conference on Ubiquitous computing&2008\\
S45&Fish’n’Steps: Encouraging Physical Activity with an Interactive Computer Game&Lin, James J., Mamykina, Lena, Lindtner, Silvia, Delajoux, Gregory, Strub, Henry B.&International Conference on Ubiquitous Computing&2006\\
S46&Study of the Usability of an Adaptive Smart Home Interface for People with Alzheimer’s Disease&Gullà, Francesca, Menghi, Roberto, Germani, Michele&Italian Forum of Ambient Assisted Living&2019\\
S47&Combined Health Monitoring and Emergency Management through Android Based Mobile Device for Elderly People&Kozlovszky, Miklos, Sicz-Mesziár, János, Ferenczi, János, Márton, Judit, Windisch, Gergely, Kozlovszky, Viktor, Kotcauer, Péter, Boruzs, Anikó, Bogdanov, Pál, Meixner, Zsolt, Karóczkai, Krisztián, Ács, Sándor&International Conference on Wireless Mobile Communication and Healthcare&2012\\
S48&Adapting Web-Based Information to the Needs of Patients with Cancer&Diana Bental, Alison Cawsey, Janne Pearson, Ray Jones&International Conference on Adaptive Hypermedia and Adaptive Web-Based Systems&2000\\

\label{tab:listofstudies}
    \end{longtable}%
%\end{table}%
}
\section{Glossary}

AUI  Adaptive User Interface
SLR  Systematic Literature Review\\
RFCD Risk Factors Corresponding to Chronic Diseases\\
AAL  Ambient Assisted Living\\
UI   User Interface\\
RQ   Research Question\\
SE   Software Engineering\\
NCD  Non-Communicable Disease\\
AR   Augmented Reality\\
VR   Virtual Reality \\
QA   Quality assessment\\
HealPM Health Promoting and Self-monitoring\\
HIM  Healthcare information management\\
CNNs Convolutional Neural Networks\\

\end{appendices} 

\bibliography{sn-bibliography}% common bib file


\end{document}